\documentclass[useAMS,usenatbib,a4paper]{mn2e}

\usepackage[totalwidth=500pt,totalheight=710pt]{geometry}
\usepackage{xspace}
\usepackage{graphicx}
\usepackage{amssymb}
\usepackage{amsmath}

\newcommand{\apj}{ApJ}
\newcommand{\apjl}{ApJL}

\newcommand{\mnras}{MNRAS}

\newcommand{\aap}{A\&A}
\newcommand{\aj}{AJ}
\newcommand{\prd}{Phys. Rev. D}
\newcommand{\nat}{Nature}

\newcommand{\etal}{et~al.~}
\def\spose#1{\hbox  to 0pt{#1\hss}}  
\newcommand{\lta}{\mathrel{\spose{\lower 3pt\hbox{$\sim$}}\raise  2.0pt\hbox{$<$}}}
\newcommand{\gta}{\mathrel{\spose{\lower  3pt\hbox{$\sim$}}\raise 2.0pt\hbox{$>$}}}
\newcommand{\eg}{{\it e.g.\ }}
\newcommand{\ie}{{\it i.e.\ }}
\newcommand{\be}{\begin{equation}}
\newcommand{\ee}{\end{equation}}
\newcommand{\dee}{\, \mathrm{d} \!}
\newcommand{\bea}{\begin{eqnarray}}
\newcommand{\eea}{\end{eqnarray}}

\def\Sref#1{Section~\ref{#1}\xspace}
\def\Fref#1{Figure~\ref{#1}\xspace}

\def\Eref#1{Equation~\ref{#1}\xspace}

\def\Aref#1{Appendix~\ref{#1}\xspace}

\newcommand{\kms}{\ifmmode  \,\rm km\,s^{-1} \else $\,\rm km\,s^{-1}  $ \fi }
\newcommand{\kpc}{\ifmmode  {\rm kpc}  \else ${\rm  kpc}$ \fi  }  
\newcommand{\pc}{\ifmmode  {\rm pc}  \else ${\rm pc}$ \fi  }  
\newcommand{\Msun}{\ifmmode {\rm M_{\odot}} \else ${\rm M_{\odot}}$ \fi} 
\newcommand{\Zsun}{\ifmmode {\rm Z_{\odot}} \else ${\rm Z_{\odot}}$ \fi} 
\newcommand{\yr}{\ifmmode yr^{-1} \else $yr^{-1}$ \fi} 
\newcommand{\hMsun}{\ifmmode h^{-1}\,\rm M_{\odot} \else $h^{-1}\,\rm M_{\odot}$ \fi}

\newcommand{\MS}{Millennium Simulation\xspace}

\def\zd{z_{\rm d}}
\def\zs{z_{\rm s}}

\def\Dt{D_{\Delta t}}

\def\xkappa{\kappa_{\rm ext}}
\def\kappax{\kappa_{\rm ext}}
\def\kappaxtrue{\kappax^{\rm true}}
\def\kappah{\kappa_{\rm h}}

\def\kappatrue{\kappa_{\rm true}}

\def\kappah{\kappa_{\rm h}}

\def\gammax{\gamma_{\rm ext}}

\def\Mstar{M^{*}}

\def\Mstarobs{\Mstar_{\rm obs}}

\def\Mhalo{M_{200}}

\def\pr{{\rm P}}
\def\data{{\mathcal{D}}}

\usepackage[usenames]{color}

\newcommand{\comment}[1]{}
\newcommand{\comments}[1]{}

\newcommand{\new}[1]{{#1}}

\newcommand{\phil}[1]{#1}
\newcommand{\tom}[1]{#1}

\def\devided{divided\xspace}

\def\infered{inferred\xspace}

\def\dependant{dependent\xspace}
\def\proceedure{procedure\xspace}
\def\propogate{propagate\xspace}
\def\propogates{propagates\xspace}

\def\ioa{Institute of Astronomy, University of Cambridge,
  Madingley Rd, Cambridge, CB3 0HA, UK}
\def\oxford{Dept.\ of Physics, University of Oxford, 
  Keble Road, Oxford, OX1 3RH, UK}
\def\kipac{Kavli Institute for Particle Astrophysics and Cosmology, 
 Stanford University, 452 Lomita Mall, Stanford, CA 94035, USA}
\def\ucsb{Dept.\ of Physics, University of California, 
  Santa Barbara, CA 93106, USA}
\def\davis{Dept.\ of Physics, U.C.~Davis, Davis, CA 95616, USA}
\def\kapteyn{Kapteyn Astronomical Institute, University of Groningen, 
  P.O.Box 800, 9700 AV Groningen, The Netherlands}
\def\asiaa{Institute of Astronomy and Astrophysics, Academia Sinica, P.O.~Box 23-141, Taipei 10617, Taiwan}

\def\collettemail{\tt t.collett@ast.cam.ac.uk}

\def\packard{Packard Research Fellow}

\title[Line of Sight Mass Reconstruction]
{Reconstructing the Lensing Mass in the Universe \\
from Photometric Catalogue Data}
    
\author[Collett \etal]{%
  Thomas~E.~Collett$^{1}$\thanks{\collettemail},
  Philip~J.~Marshall$^{2}$,
  Matthew~W.~Auger$^{1}$,
  Stefan~Hilbert$^{3}$,
\newauthor{%
  Sherry~H.~Suyu$^{4,3,5}$,
  Zachary~Greene$^{4}$,
  Tommaso~Treu$^{4}$\thanks{\packard},
  Christopher~D.~Fassnacht$^{6}$,}
\newauthor{%
  L\'eon~V.~E.~Koopmans$^{7}$,
  Maru\v{s}a Brada\v{c}$^{6}$,
  Roger~D.~Blandford$^{3}$} 
  \medskip\\
  $^1$\ioa\\
  $^2$\oxford\\
  $^3$\kipac\\
  $^4$\ucsb\\
  $^5$\asiaa\\
  $^6$\davis\\
  $^7$\kapteyn
}

\begin{document}
             
\date{Accepted for publication in MNRAS}
\pagerange{\pageref{firstpage}--\pageref{lastpage}}\pubyear{2012}

\maketitle           

\label{firstpage}


\begin{abstract} 

High precision cosmological distance measurements towards individual objects
such as time delay gravitational lenses or type Ia supernovae are affected by
weak lensing perturbations by galaxies and groups along the line of sight. In
time delay gravitational lenses, ``external convergence,'' $\kappax$, can
dominate the uncertainty in the inferred distances and hence cosmological parameters.
 In this paper we attempt
to reconstruct $\kappax$, due to line of sight structure, using a simple halo
model.
We use mock catalogues from the Millennium Simulation, and calibrate and 
compare our reconstructed $\pr(\kappax)$ to ray-traced $\kappax$ ``truth''
values; taking into account realistic uncertainties on redshift and stellar
masses. \tom{We find that the reconstruction of $\kappax$ provides an improvement in
precision of $\sim$50\% over galaxy number counts.} 
We find that the lowest-$\kappax$ lines of sight have the best constrained 
$\pr(\kappax)$. In anticipation of future samples with thousands of lenses,
we find that selecting the third of the systems with the
highest precision $\kappax$ estimates gives a  sub-sample of unbiased time
delay distance measurements \phil{with (on average) just} 1\% uncertainty due
to line of sight external convergence effects. Photometric data alone are
sufficient to pre-select the best-constrained lines of sight, and can be done
before investment in light-curve monitoring. 
\phil{Conversely, we show that selecting lines of sight
with high external shear could, with the reconstruction model presented here,
induce biases of up to 1\% in time delay distance.} We find that a major potential source of systematic error is uncertainty in the high mass end of the stellar mass-halo mass relation; this could introduce $\sim$2\% biases on the time-delay distance if completely ignored. We suggest areas for the
improvement of this general analysis framework (including more sophisticated
treatment of high mass structures) that should allow yet more accurate
cosmological inferences to be made.

\end{abstract}



\setcounter{footnote}{1}


\section{Introduction}
\label{sec:intro}

Every distant object we observe has had its apparent shape distorted,
and size and total brightness magnified (or demagnified) by a compound
weak gravitational lens constructed from all the mass distributed
between us and it. As \citet{Vale+White2003} and \citet{HilbertEtal2007}
showed, there are no empty lines of sight through our universe. This
fact makes gravitational lensing a potentially important source of
systematic error for any estimate of luminosity (or distance); this 
issue has been 
raised for \eg type Ia supernovae by
\citet[][]{Holz+Wald1998,Holz+Linder2005}, for gamma ray bursts by
\citet[][]{Oguri+Takahashi2006,Wang+Dai2011},  and for high redshift
galaxies by \citet{WyitheEtal2011}, among others. 

Along lines of sight containing strong gravitational lenses, the perturbative
effects of line of sight mass structure have been found to be particularly
important, with foreground and background structures having a significant
effect on the inferred lensing cross-section \citep[\eg][]{WongEtal2012} and
distance ratios \citep[][]{DalalEtal2005}. Indeed, \citet{SuyuEtal2010} found
that in time delay lens cosmography the so-called ``external convergence''
$\kappax$ due to mass structures along the (unusually over-dense) line of
sight to the quadruply-imaged radio source B1608$+$656 had to be included in
their analysis, and was the dominant source of uncertainty in their 5\%
measurement of the time  delay distance.  Large samples of time-delay lenses
are expected to be discovered in the next decade in ground-based optical
imaging surveys \citep{Oguri+Marshall2010};  the external convergence will
have to be understood increasingly well in order to prevent its correction
from dominating the systematic error budget.

While it is rare for three galaxies to line up well enough for both of
the background sources to be strongly lensed \citep{GavazziEtal2008,CollettEtal2012a}, 
the large size of dark matter halos makes partial alignments -- such that they
act as perturbing weak lenses --  a near certainty. The large scale of the
perturbers means the external lensing perturbations can be approximated
by a quadrupole lens characterised only by external convergence and shear.
The external shear, $\gammax$ can potentially be recovered in the
mass-modelling of the lens, but the convergence is undetectable from the
image positions, shapes and relative fluxes; this is the well-known {\emph{ 
mass-sheet degeneracy}} \citep[see e.g.][for details]{FalcoEtal1985}.

Since the mass-sheet degeneracy prevents any estimate of $\kappax$ from the 
strong lens modelling, additional information is required. Weak lensing measurements
can be used for rich lines of sight \citep{NakajimaEtal2009, FadelyEtal2009}, but 
for lines of sight with low galaxy density we must attempt to reconstruct the distribution
of mass along the line of sight. Attempts to reconstruct the mass distribution in strong
lens fields have focused on understanding the external shear which seems to be large in most strong lenses.
Surveys by \citet{Fassnacht+Lubin2002,AugerEtal2007,WilliamsEtal2006,MomchevaEtal2006,FassnachtEtal2006}
all found groups of galaxies hosting or near to known gravitational lenses.
Is generally found that lens galaxies, reside in over-dense environments,
indistinguishable from those occupied by similarly massive galaxies \citep{Auger2008,TreuEtal2009}
\citet{WongEtal2011} estimated $\Pr(\gammax)$ given the data of \citet{WilliamsEtal2006}
and \citet{MomchevaEtal2006} but found significant discrepancy between the predicted shear
distribution and the external shear demanded by the strong lens model. 
\citeauthor{WongEtal2011} also found that both line of
sight structures and the group of galaxies in each lens plane contribute
significant proportions of the shear. 

For lines of sight without strong lenses, magnification is the more relevant
quantity than convergence; \citet{GunnarssonEtal2006} used a galaxy halo model
combined with simple scaling relations to reconstruct supernovae lines of
sight and found that the dispersion in apparent source brightness due to
lensing magnification could be reduced by a factor of two. Conversely
\citet{KarpenkaEtal2012} used the brightness dispersion in type Ia supernovae
to infer the parameters of both their galaxy halo model and mass-to-light
scaling relation. Both \citeauthor{KarpenkaEtal2012} and
\citet{JonssonEtal2010} were able to  detect lensing effects in the supernova
data this way.

Large numerical simulations can also be used to estimate global convergence
distributions. \citet{Holder+Schechter2003} and \citet{DalalEtal2005} carried
out ray-tracing calculations in N-body simulations \citep{KauffmannEtal1999,WambsganssEtal2004} to estimate the
distribution of external shear values.  \citet{HilbertEtal2009} performed
similar ray-tracing experiments through the Millennium Simulation
\citep{SpringelEtal2005}, generating a predicted $\kappax$ at every position
in a simulated sky. \citet{HilbertEtal2009} found that after removing the
matter on the strong lensing plane \MS lines of sight with strong lenses were
not biased towards high $\kappax$, although selection functions for
discovering samples of strong lenses were not taken into account.

The \MS results have been used to analyse real observations as well. 
\citet{SuyuEtal2010} selected \MS lines of sight by their apparent
galaxy over-density, in a 45 arcsec radius aperture down to $i < 24.5$,
to match the observed over-density towards the time delay lens
B1608$+$656 \citep{FassnachtEtal2011}.  The resulting distribution of
$\kappax$ values from the ray-tracing was taken to be an estimate of 
$\Pr(\kappax)$, which was then marginalised over when inferring the time
delay distance in this system. A similar procedure was used in \citet{SuyuEtal2012}.
In a companion paper to this one, \citet{Greene} investigate
improvements to this method by weighting the galaxy counts by observables
including galaxy luminosity and perpendicular distance to the line of sight, again using the
\MS mock catalogues and their associated $\kappax$ values to construct
$\Pr(\kappax|\data)$. For the most over-dense lines of sight their
results  show a $\sim$20\% improvement over
number counts alone, but little improvement for less dense ones.

\begin{figure*}
\includegraphics[width=\textwidth]{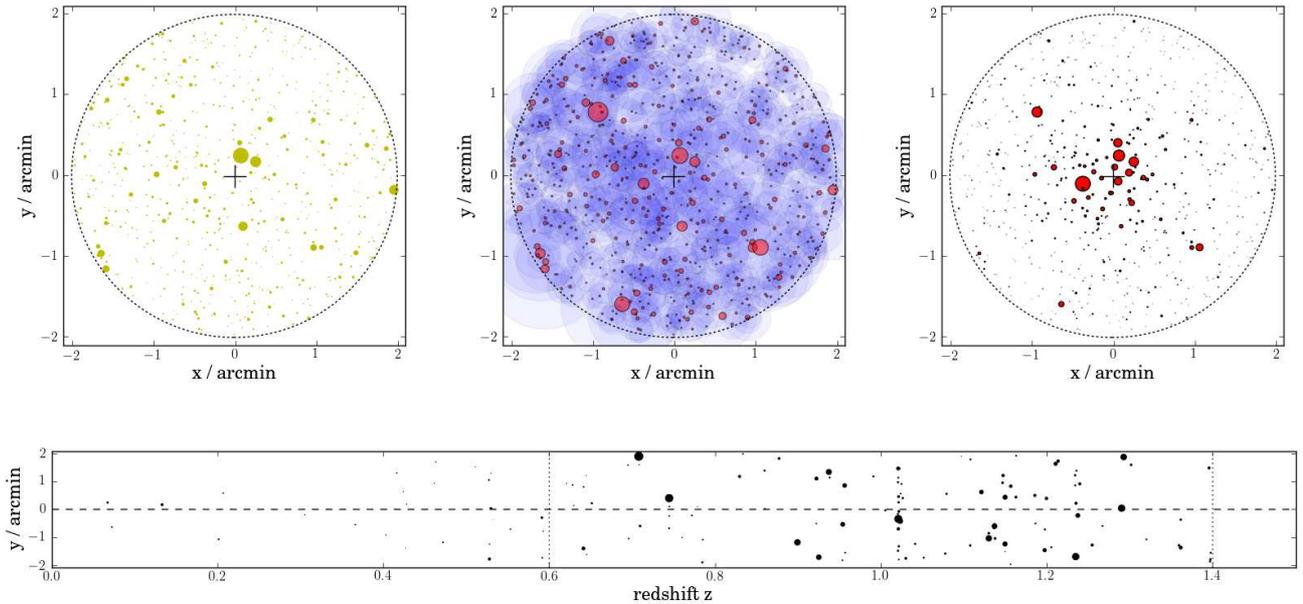}
\caption[magcut]{Four different views of a slightly over-dense \MS line
of sight, with a $\kappax$ of 0.03. 
{\it Top row, left:} The positions of galaxies projected on the sky. The area
of the circles is proportional to observed $i$-band flux; the brightest
object shown has an $i$ magnitude of 17.7. 
{\it Top row, centre:} The angular sizes of halos projected on the sky.
Red and blue regions lie within the NFW scale radius and virial radius
of each halo, respectively. There are essentially no empty lightcones. 
{\it Top row, right:} The individual $\kappax$ contributions of
each halo, assuming a \citet{BMO} truncated NFW 
profile and the \citet{Neto2007}
mass-concentration relation. Comparison of
this panel and the centre panel illustrates the relative importance of
proximity to the line of sight.
{\it Bottom:} A view along the redshift axis, showing only halos with
$|x|<0.3$ arcmin. The area of the points is proportional to each halo's
mass: the most massive halo shown has $1.6\times10^{12}\Msun$. The
optical axis is shown by the dashed line, while the dotted lines mark
the lens and source planes for a B1608-like strong lens.}
\label{fig:lightcone}
\end{figure*}

In this paper, we combine the halo model reconstruction approach of
\citet{GunnarssonEtal2006}, \citet{JonssonEtal2010}, 
\citet{WongEtal2011} and others, with the
idea of calibrating to simulations from \citet{SuyuEtal2010}
Taking the $\kappax$ values from
\citet{HilbertEtal2009}'s ray-tracing calculations as the ``truth''
that we have to recover, we use the \MS mock galaxy catalogues first to
{\it calibrate} the reconstruction and account for unseen mass and
voids, and then to {\it test} the accuracy of the line of sight mass
reconstruction under this assumption. Probabilistically assigning mass
and redshift to every observed galaxy in a given observed field, we
generate Monte Carlo sample line of sight mass distributions, and so
construct the probability distribution function (PDF) $\Pr(\kappax|\data)$ for that field.
We then emulate
the combination of many such PDFs to quantify the residual bias that
would be translated to the global (including cosmological) parameters, 
\phil{if left unaccounted for.}
In doing so, we aim to answer the following questions: 

\begin{itemize}

\item Faint galaxies, filaments and dark structures will not appear in
any photometric object catalogue, but they will contribute convergence at
some level. How much of the total line of sight 
convergence comes from visible
galaxies, and how much effect do dark structures and voids have? 

\item Can the true line of sight convergence be recovered from a calibrated
halo model reconstruction? What scatter and residual bias are induced by the
reconstruction process, and how might these be reduced in the future? 

\item Both the detectability of lenses, and the selection of sub-samples of
them for further observation, could potentially induce a selection bias in
external convergence. How should we select objects to achieve the highest
accuracy convergence reconstruction? Can such a selection be made robustly,
using the reconstruction results?

\end{itemize}

This paper is organized as follows. We review the relevant gravitational
lensing theory in~\Sref{sec:theory}, and the Millennium Simulation
ray-traced convergences and mock catalogues in \Sref{sec:MS}, before
introducing our simple reconstruction  model in \Sref{sec:model}. We
then test this model in two phases: first, in \Sref{sec:knownMh+z}, with
known redshift and halo mass for every galaxy in a lens field, in order
to quantify the irreducible uncertainty due to unseen mass, and second,
in \Sref{sec:obsMstar+z}, with realistic observational uncertainties on
the observed galaxies' stellar masses and redshifts. In
\Sref{sec:biases} we investigate the potential systematic error induced
by selecting a subset of lines of sight. In
\Sref{sec:SHAMfail} we investigate systematic errors caused by assuming an incorrect stellar-mass relation. We discuss our results in
\Sref{sec:discuss} before concluding in \Sref{sec:conclude}.

Throughout this paper magnitudes are given in the AB system and
we adopt the Millennium Simulation's ``concordance'' parameters for our reference cosmology, \ie
$h=0.73$, $\Omega_m=0.25$ and $\Omega_\Lambda=0.75$, where the symbols indicate
the Hubble Constant in units of 100 km s$^{-1}$ Mpc$^{-1}$ and the matter and
dark energy density of the Universe in units of the critical density.

\section{Theoretical Background}
\label{sec:theory}

In strong lens systems where the source is time variable, the images do not 
vary simultaneously; the optical path length for each image is different due
to relativistic and geometric effects. The difference in optical path length
causes a time-delay between the light curves of each image.
\citet{FalcoEtal1985} showed that the presence of additional matter (in the
form of a mass-sheet) along the line of sight has no observable effect except
to rescale the time-delays and magnifications. As such, it is necessary to include $\kappax$ in
the lens modelling if cosmological parameters are to be estimated accurately
and precisely from observed time delays \citep{SuyuEtal2010}.

If there is external convergence present that is not included in the
lens modelling, then the time delay distance -- inferred assuming $\kappax
= 0$ -- will be $(1-\kappax)$ more than the true value of the time-delay distance $\Dt$:
\be 
\label{eq:MassSheet:Dtbias}
\Dt^{\rm{true}}=\frac{\Dt^{{\kappax = 0}}}{1-\kappax}.
\ee

We can hence estimate the true distance {\it only} if we have additional
knowledge of $\kappax$. Since $\kappax$ is typically small the absolute
uncertainty on the estimate of $\kappax$ corresponds to the fractional
uncertainty with which time-delay distances can be \infered.

Dynamical observations of the lens galaxy
can help break the mass-sheet degeneracy by providing an additional
estimate of the lens's mass
\citep[e.g.,][]{KoopmansEtal2003,Koopmans2004,SuyuEtal2010}. This constraint 
is useful in excluding the high convergence tails that are often present in 
PDFs for convergence - we do not include dynamical constraints in our analysis. 

In this paper we restrict ourselves to reconstructing
the line of sight portion of the mass distribution in any given field, {\it our lines of sight do not include a strong lens} which allows us to work in the weak lensing regime.

The impact of weak-lensing perturbations on strong lensing lines-of-sight is
postponed to further work.


\section{The Millennium Simulation}
\label{sec:MS}

In order to test the accuracy of our convergence estimates, we need to
know the true convergence for each line of sight. We cannot use  real
lines of sight for this,  but must use simulated lines of sight instead.
In this section we briefly review the Millennium Simulation, the ray
tracing calculations that have been carried out in it, and the mock
galaxy catalogues that have been produced.

The \MS \citep{SpringelEtal2005} is a cosmological N-body simulation of
Dark Matter structures in a cubic region approximately 680~Mpc in
co-moving size, followed from redshift 127 to the present day. With an
approximate halo mass resolution of $2\times10^{10}{\rm M}_{\odot}$
(corresponding approximately to a galaxy with luminosity $0.1L^{*}$), it
provides a detailed prediction for the distribution of dark structures
present in the Universe, under the assumptions of the $\Lambda$CDM model
of hierarchical structure formation, cosmological densities as given at
the end of \Sref{sec:intro}, and $\sigma_8 = 0.9$.
Galaxy properties,
such as stellar masses, luminosities and colours, were assigned to the
simulated halos according to a semi-analytic model for galaxy formation
\citep{DeLucia+Blaizot2007}: the resulting predictions for the galaxy
luminosity function and correlation function match the observational
data very well.


\subsection{Convergence from Ray Tracing}
\label{sec:MS:raytracing}

\citet{HilbertEtal2009} calculated the lensing convergence using the
``lightcone'' dark matter only  output of the \MS, providing 1.7 degree
square  mock sky maps of this quantity. In this process, the dark matter
density from the simulation was projected onto a set of lens planes at
discrete redshifts, adaptively gridded and smoothed, and then the second
derivatives of the  lensing potential required for the components of the
magnification matrix were computed using a multiple lens plane ray tracing
algorithm. The first order approximation to this algorithm \citep[equation 17
of][]{HilbertEtal2009} gives the total convergence at a given sky
position~$\bmath{\theta}$ as the simple summation of the surface density in
each lens plane, weighted by the inverse critical surface density $\Sigma_{\rm
cr}$ for that plane's redshift:
\be
\kappax(\bmath{\theta}) = \sum_i \frac{\Sigma_i(\bmath{\theta})}
                                     {\Sigma_{\rm cr}(z_i,\zs)}.
\ee
\citet{HilbertEtal2009} found that this approximation is accurate to a percent level
in $\kappax$, even on 30 arcsecond scales. This accuracy sufficient for our analysis, but 
the approximation may need re-visiting in future work.

To define a test line of sight, we draw a random sky position from
within the convergence maps, bi-linearly interpolate between the pixels
of the map, and store the result as the ``true'' convergence for that
line of sight, $\kappax^{\rm true}$.

\new{The adaptive smoothing of the matter in the \MS results in a spatial resolution of $\sim 5\,\text{kpc}$ comoving in the densest regions, and $\sim10\,\text{kpc}$ on average. This resolution is sufficient to capture most of the variance in the convergence \citep{TakahashiEtal2011}. Moreover, we seek to estimate the component of the convergence that is smooth on the angular scales of typical strong galaxy-scale lens systems ($\sim 1\,\text{arcsec}$). Any variations of the external convergence on smaller scales can be directly recovered from the resulting image distortions in the lens modelling.}

\comment{PJM: This means our sightlines are not typical of strong lenses, because they
don't include the local overdense region typically found near massive lens
galaxies. Our lines of sight have typical mass distributions for a lens line
of sight {\it after excision of the lens plane}. I have tried to explain
this above.}


\subsection{Mock Photometric Galaxy Catalogues}
\label{sec:MS:mocks}

We construct mock photometric galaxy catalogues by selecting all \MS galaxies
within a circular aperture of a given angular radius centred on each randomly
selected line of sight position, with apparent $i$-band magnitude brighter
than a given limit. The parent catalogue, from \citet{HilbertEtal2011},
contains galaxy positions and magnitudes, all of which include lensing effects
(deflections and magnifications). We also have access to the underlying galaxy
halo masses and redshifts, which we will use to calibrate the reconstruction,
and also to explore any sources of bias and scatter. Since the ``true''
convergence is that of only the dark matter, we focus on reconstructing the
dark matter halos alone. Our model for doing this is outlined in the next
section. 


\section{Estimating Convergence: The Halo Model Approximation}
\label{sec:model}

We require a method for estimating the convergence at any point on the sky,
given a catalogue of observed galaxy properties such as positions,
brightnesses, colours, and possibly stellar masses and redshifts, for every
galaxy in the field of the lens, down to some magnitude limit. Since we cannot
observe the surface mass density of each halo directly, we need some way of
estimating the mass of each halo in the catalogue, so that we can compute its
contribution to the total $\kappax$ along the line of sight \citep[as in
\eg][]{GunnarssonEtal2006,KarpenkaEtal2012}.  This mass
assignment recipe will be uncertain, and also incomplete, but it can be
calibrated with cosmological simulations, which represent a significant
additional source of information to what is present in the data.

Transforming an observed photometric galaxy catalogue into a  catalogue of
halo masses, positions, and redshifts will enable us to attempt a
reconstruction of the convergence induced by every halo near a line of sight. 
We also need to account for the convergence due to dark structures and the
divergence due to voids; we do not model voids or dark structures, but their
effects are included statistically by calibrating the convergence in
halos, $\kappah$, against the ray-traced convergence, $\kappax$, for \MS lines
of sight.


\subsection{Halos}
\label{sec:model:halos}

Cosmological dark matter simulations have shown that dark matter
halos are reasonably well-approximated by NFW profiles 
\citep{NFW}. We assume each halo to have a spherical mass distribution with 
density given by
\be\label{eq:rhonfw}
\rho(r) = 
\frac{\rho_0}{\left(r/r_{s}\right)
\left(1+r/r_{s}\right)^{2}}.
\ee
Here, $r_{s}$ is a characteristic scale 
radius of the cluster, representing the point where
the density slope transitions from $r^{-1}$ to $r^{-3}$. This radius is
related to the virial radius of the halo by $r_{s}~=~r_{200}/c$, where $c$ is
the concentration parameter, which can be estimated from the halo's mass,
using a mass--concentration relation. Typically more massive halos are less
concentrated, but there is some scatter; we use the relation of
\citet{Neto2007} to estimate $c$ from the mass enclosed within $r_{200}$,
which we denote as $M_{200}$. We find our results do not change if the
\citet{MaccioEtal2008} mass--concentration relation is used instead. 

If one integrates the density profile of an NFW profile out to infinite
radius, the total mass diverges. Similarly, if the universe is
homogeneously populated with NFW halos, the projected surface mass along
any line of sight will also be divergent.\footnote{At large radius,
$\Sigma_{\rm nfw} \propto R^{-2}$, but the differential number of halos
centred within an annulus of width $\dee R$ is given by $\dee N_{\rm
annulus} \propto R \dee R$, so  $\Sigma_{\rm total} \propto
\int_{0}^{\infty} R^{-1} \dee R$, which diverges logarithmically} Since
infinite mass is unphysical, the profile must be truncated at some
point. Several truncation profiles have been suggested
\citep[e.g][]{BMO}, but beyond several virial radii, the amount of
matter associated with a halo is likely to be low. In this work we
assume the truncated NFW profile
\be\label{eq:bmoprofile}
\rho(r) = 
\frac{\rho_{\rm NFW} (r)}{1+\left(r/r_{t}\right)^2},
\ee
which is the same as the NFW profile in the limit that the truncation
radius, $r_t$ goes to infinity; the shear and convergence from such a
profile are derived in \citet{BMO}.
 We use a truncation radius of five times
the virial radius, but our results are robust for any choice of $r_t>2
\times r_{200}$. 

Many studies have shown that galaxies have total mass density profiles that
are approximately isothermal in their inner regions
\citep[\eg][]{AugerEtal2010} due to the more concentrated stellar mass
component.
This has been confirmed by galaxy-galaxy weak lensing measurements 
\citep[\eg][]{Mandelbaum,GavazziEtal2007}. We note that the two halo term that
dominates the galaxy-galaxy lensing signal at large radii is explicitly taken
into account in our model, where every galaxy is assigned a dark matter halo.
At large radii an NFW-like profile is probably appropriate, but may not be
correct when a halo is very close to the line of sight.

Each halo in our catalogue contributes to
the line of sight convergence by
\be
\label{eq:kappai}
\kappa_i =\Sigma_{i}/\Sigma_{\mathrm{cr}}(z_i,\zs),
\ee
where $\Sigma_{\mathrm{cr}}(z_i,\zs)$ is the critical surface density \comments{was defined in \Eref{eq:sigcrit}}.
Following the first order approximation of \citet{HilbertEtal2009}
outlined in \Sref{sec:MS} above, we compute 
the total convergence from all the halos along the line
of sight using
\be 
\label{eq:kappasummu}
\kappah = \sum_{i} \kappa_i.
\ee

To estimate a halo's contribution to $\kappah$ we must first know the
halo's mass \new{and position. The stellar mass
and host halo are not necessarily concentric, especially for central
galaxies of giant clusters, but our reconstruction assumes the dark matter halo is centred on
the visible galaxy.} For the lines of sight of the \MS, halo masses are
known, but for  real lines of sight the halo masses have to be \infered
from whatever data are available. 
In this work we investigate the use of the empirical stellar mass --
halo mass relation for this purpose. The primary advantage of this
approach is that it is a simple, ``one size fits all'' relation that can
be applied regardless of the stellar mass observed. We discuss
additional observables that could be used to improve the precision of
the halo mass inference in \Sref{sec:discuss} below. 

We use the relation of \citet{BehrooziEtal2010} to infer halo masses
from stellar masses; the details of this \proceedure are outlined in 
\Aref{appendix:MSMH}. Given an uncertain stellar mass and redshift for
each galaxy in the  catalogue, we draw a sample halo mass from the PDF
that describes this uncertainty (\Eref{eq:mhalo-mstar}) and use
\Eref{eq:kappai} to compute its contribution to the convergence,
$\kappa_i$; this can be done for all the halos along a specific line of sight
and summed to give a sample value of $\kappah$;  repeatedly applying
this \proceedure allows us to build up a histogram of $\kappah$ values
consistent with the data, and hence characterize $\pr(\kappah|\data)$. 

This PDF contains a hidden assumption of an uninformative prior PDF for
$\kappah$ -- in the case of infinitely poorly measured stellar masses and
redshifts, $\pr(\kappah|\data)$ will have very long tails corresponding to
very over-dense lines of sight that we do not believe exist. The resolution of this
is to divide out the effective prior PDF that was applied during the halo mass
estimation process (which is broad, and hence close to uniform), and apply an
additional prior on $\kappah$, given by the global underlying $\kappah$
distribution from the simulation. In the limiting  case of very poor
photometric data, $\pr(\kappah|\data)$ then defaults to this distribution, as
required.




\subsection{Accounting for voids, filaments and other dark structures}
\label{sec:model:voids}

Our halo model accounts for the convergence contribution of density
perturbations that can be associated with light. Filaments and dark
substructures contribute additional convergence, while voids contribute a
divergence; these structures must be included to produce an unbiased estimate
of $\kappax$. In particular, neglecting voids would lead to a heavily biased
estimate of $\kappax$, since the halo model's $\kappah$ can only be positive.
In principle the absence of galaxies implies something about the presence of a
void, but we do not currently attempt to use this information. In general, it
is difficult to account for the unseen mass, since we do not have a good model
for its density structure.

The solution we present here is to calibrate the relationship between the halo
convergence $\kappah$ and the overall convergence $\kappax$ (as would be
calculated by ray tracing through the full density field)  using the
simulations themselves. The halo modelling procedure outlined in
\Sref{sec:model:halos} allows  us to estimate  the convergence due to halos
given observations of  mass, redshift and position,
$\pr(\kappah|\data)$, but what we are interested in is the total
convergence along the line of sight $\pr(\kappax|\data)$. We can obtain
this by considering the expression: 
\begin{equation} 
\Pr(\kappax|\data) = 
   \int \dee\kappah  \Pr(\kappax|\kappah,\data) \Pr(\kappah|\data)
   \label{eq:kappaconv}
\end{equation} 
The first term in the integrand relates the convergence due to model halos,
$\kappah$, to the true convergence, $\kappax$. This conditional distribution
can be constructed from the \MS catalogues, by computing
$\kappah$ from the true halo masses and redshifts 
along each selected line of sight, and then
accumulating $\kappah$, $\kappax$ pairs
for a large number of such sightlines. 
For any given line of sight we can then estimate
$\pr(\kappah|\data)$ as described above, and then multiply it by
$\Pr(\kappax|\kappah)$ and integrate out the intermediate parameter
$\kappah$. 

\phil{However, if the conversion between stellar mass and halo mass is very
uncertain, $\pr(\kappah|\data)$ is shifted relative to what it would
have been given perfect knowledge of the halo masses.  This is because the
conversion from stellar mass to halo mass and into $\kappah$ is highly
asymmetric: a long tail of high halo masses gives rise to a tendency to
overestimate $\kappah$. If this shift is ignored, the resulting
$\Pr(\kappax|\data)$ will be systematically biased towards high values of
$\kappax$. Instead, we need the  distribution of $\kappax$ from all
calibration lines of sight that have a $\pr(\kappah|\data)$ that is identical
to the $\pr(\kappah|\data)$ for the real line of sight {\it given the same
data quality}; however, we found this approach infeasible  given our finite
number of calibration lines of sight. A working compromise is to use 
the median of the $\pr(\kappah|\data)$ distributions in the calibration.}

We  take a large number of
calibration lines of sight from the \MS catalogues and generate a mock
stellar mass catalogue for each one. We then reconstruct each lightcone in the
same way as we would an observed field,
estimate $\pr(\kappah|\data)$, and extract its median. Accumulating
these median values $\kappah^{\rm med}$ and 
their corresponding true values of $\kappax$, we can form the 
conditional distribution $\pr(\kappax|\kappah^{\rm med})$. 
Then, to infer $\kappax$ from an observed photometric catalogue, we 
estimate $\pr(\kappah|\data)$ 
using the halo model, compute the median value 
$\kappa_{\rm h,obs}^{\rm med}$, and then use a modified
version of 
\Eref{eq:kappaconv} to infer $\kappax$: 
\bea
\Pr(\kappax|\kappah^{\rm med},\data) &=& \int \dee\kappah^{\rm med} 
   \Pr(\kappax|\kappah^{\rm med},\data) \Pr(\kappah^{\rm med}|\data) \notag \\
\label{eq:calkappaconv}   
\eea
\phil{This integral is trivial, since $\Pr(\kappah^{\rm med}|\data)$ is
a delta function centred on the median of the inferred PDF for $\kappah$.
However,}
because we only have a finite number of calibration lines of sight, we use all
calibration lines of sight with $\kappa_{\rm h,cal}^{\rm med} =
\kappah^{\rm med}\pm0.003$ to form $\Pr(\kappax|\kappah^{\rm
med},\data)$.

This procedure of reducing the full PDF for $\kappah$ to its median increases
the importance of the realism of the simulation: the simulation needs to be as
similar as possible to the real universe or there is a possibility of
systematic errors. Conversely, making use of all the information
in the photometric catalogue yields $\kappax$ estimates of increased
precision.


\section{Results: Perfect Halo Data}
\label{sec:knownMh+z} 

We now put the reconstruction \proceedure outlined above into practice, first
inferring $\kappax$ from $\kappah$ in the case of hypothetical galaxy
catalogues with noiseless redshifts and halo masses. The reason for doing this
is to check the validity of the calibration process, and then also to 
investigate the primary sources of the convergence. It will also provide us
with a measure of the intrinsic uncertainty introduced by our assumptions of
all illuminated halos being spherically-symmetric truncated NFW halos following the
Neto et al concentration--mass relation \new{and of visible galaxies being concentric with their host halos}.


\subsection{How is $\kappah$ related to $\kappax$?}

\begin{figure}
\includegraphics[width=\columnwidth]{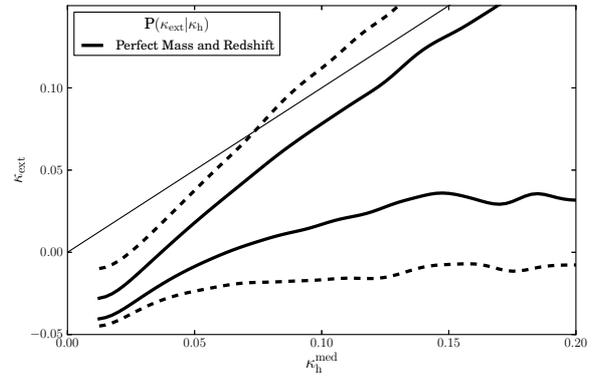}
\caption[Biased?]{The conditional distribution of 
$\kappax$ given $\kappah$, in the case where  
we have perfect knowledge of halo mass and redshift.
$10^5$ reconstructed lines of
sight were used to make this plot. 
Solid (dashed) lines enclose 68\% (95\%) of the conditional probability.
$\kappah$ traces $\kappax$, but with a positive offset which arises from
$\kappah$ not accounting for any voids or dark structures.
At fixed $\kappah$ the scatter in $\kappax$ grows rapidly with $\kappah$.
\tom{The thin diagonal line follows $\kappax=\kappah$.}}
\label{fig:jointkh-k}
\end{figure}

\Fref{fig:jointkh-k} shows the conditional distribution of $\kappax$ given 
$\kappah$, derived from $10^5$ randomly selected \MS lines of sight
and assuming noise-free halo masses and redshifts. 
\phil{In this case, $\pr(\kappah|\data)$ is almost a delta function since the only uncertainty is halo
concentration, which has minimal effect on $\pr(\kappah|\data)$.  
}
At fixed $\kappah$ we find in \Fref{fig:jointkh-k} that the scatter in 
$\kappax$
grows with $\kappah$; our reconstruction is better at reproducing under-dense
lines of sight than over-dense lines of sight. The effect of ignoring voids
and the smooth mass component is evident in this plot: $\kappah$ is
significantly higher than $\kappax$, at any given $\kappax$ value. 
\phil{Conditioning on $\kappah$ gives a PDF for
$\kappax$ centred at the correct place, by construction.}

\subsection{How accurate is the $\kappah - \kappax$ calibration?}

\begin{figure}
\includegraphics[width=\columnwidth]{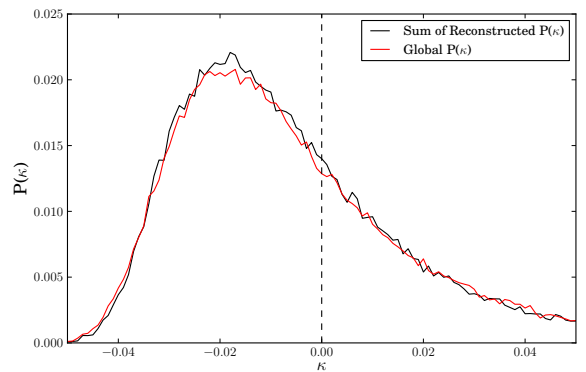}
\caption[magcut]{Recovering the global $\pr(\kappax)$ (shown in red). The
black curve shows a histogram of  inferred  $\kappax$ values from $10^{5}$
reconstructed lines of sight.  The reconstructions were performed given
perfect knowledge of halo mass and redshift -- in this case the reconstruction
recovers the correct convergence.}
\label{fig:globaldist}
\end{figure}

\begin{figure*}
\includegraphics[width=\textwidth]{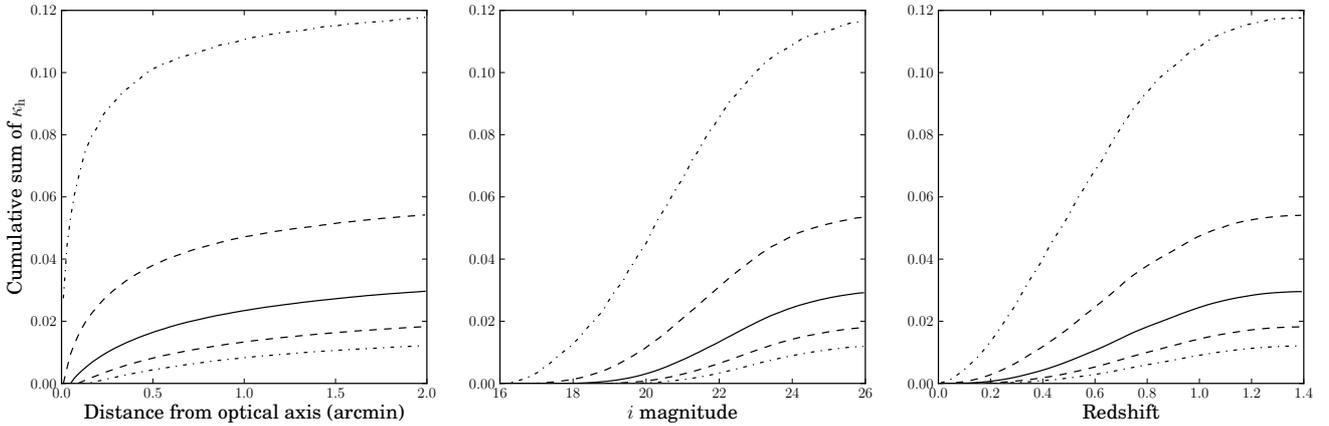}
\caption[magcut]{Which objects dominate the external convergence for a line of
sight? From left to right, the three figures show the cumulative contribution
to the total convergence from individual halos ($\kappah$) as a function of
1) distance from the line of sight, 2) magnitude and 3) redshift.   In each
panel, the solid line is the median cumulative contribution over a large
sample of lines of sight, while dashed and dot-dashed show the ranges enclosing 68
and 95 percent of the lines of sight respectively. The source is at redshift 1.4.}
\label{fig:where}
\end{figure*}

Let us define the ``bias'' of the reconstruction of a given line of sight as
the difference between the expectation value of $\kappax$ over its posterior
PDF $\pr(\kappax|\data)$ and the known true value $\kappax^{\mathrm{true}}$.
Let us also define the ``width,'' $\sigma_{\kappa}$, to be half the width of
the interval containing the central $68\%$ of the posterior probability
$\pr(\kappax|\data)$. 

In an ensemble of $10^{5}$ reconstructed lines of sight, we find the mean bias
to be $-1\times 10^{-5}$, and the mean width to be 0.01. This is
1.8 times smaller than the width of the global $\pr(\kappax)$.  We find that
our inferred $\kappax$ PDFs are consistent with the global $\kappax$
distribution; \Fref{fig:globaldist} shows that sampling from the
$\pr(\kappax|\data)$ inferred from reconstructions of  $10^{5}$ lines of
sight  produces a distribution of $\kappax$ values over the sky nearly identical to
the global $\pr(\kappax)$. Given perfect knowledge of halo mass and redshift
then, the calibration \proceedure provides an unbiased estimate of $\kappax$
that is $1.8$ times more precise than using the global $\pr(\kappax)$. 
\phil{This is the maximum precision we can obtain using the current calibrated
halo model.}


\subsection{Which halos dominate the $\kappah$ distribution?}

Before investigating the impact of imperfect knowledge of halo
mass and redshift, we investigate the uncertainties induced by limits on the
magnitude of the observed galaxy sample, and on the field of view. The halos
in our catalogue are populated by galaxies and given magnitudes according to
the semi-analytic model of \citet{DeLucia+Blaizot2007}: by applying magnitude
cuts to our catalogue, we can investigate the amount of scatter caused by
unobserved halos. 

\begin{figure}
\includegraphics[width=\columnwidth]{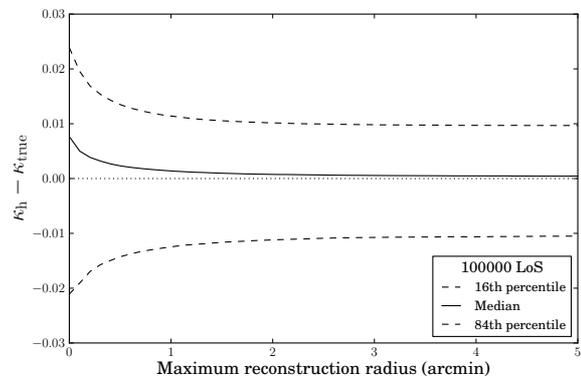}
\caption[magcut]{The 16, 50 and 84th percentiles of $\kappah$ minus
$\kappax$ as a function of the limiting radius of the halo
reconstruction. $\kappah$ has been shifted such that
$\left\langle\kappah\right\rangle=0$ \phil{to emulate the effect of the
eventual calibration}. 
The majority of the constraining power
comes from reconstructing halos within 2~arcmin of the line of sight.}
\label{fig:radcut}
\end{figure}

\begin{figure}
\includegraphics[width=\columnwidth]{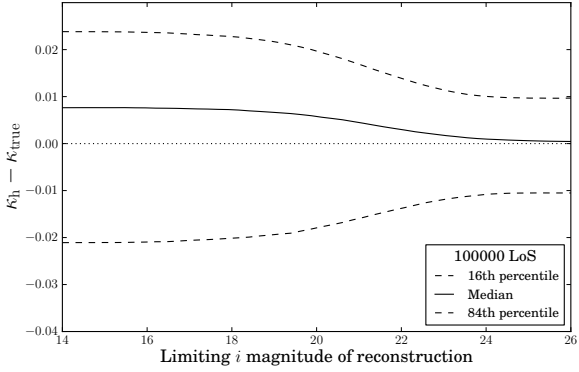}
\caption[magcut]{The 16, 50 and 84th percentiles of $\kappah$ minus
$\kappax$ as a function of the limiting $i$ band depth of the halo
reconstruction. $\kappah$ has been shifted such that
$\left\langle\kappah\right\rangle=0$ \phil{to emulate the effect of the
eventual calibration}. The majority of the constraining power
comes from reconstructing halos with magnitudes between $18<i<24$.}
\label{fig:magcut}
\end{figure}

\Fref{fig:where} shows the cumulative contribution to $\kappah$ as a function
of projected distance from the line of sight, magnitude and redshift. We find
that most of the convergence comes from halos close to the line of sight. 
Over half of the  convergence typically comes from halos within 30 arcseconds
of the line of sight; by 2 arcminutes, this fraction is 85\%. The contribution
from halos beyond 2 arcminutes is relatively constant at a contribution of
0.008$\pm$0.003. We find that ignoring halos beyond 2 arcminutes has no effect
on the precision of the reconstruction; this is shown in \Fref{fig:radcut}. As
a function of magnitude, we find that $\kappah$ is dominated by objects with
magnitudes between $i=18$ and $i=24$. Objects brighter than $i=18$ are either
too rare, or too close to the observer, to make a significant contribution to
the convergence. Objects fainter than $i=24$ are too small to be important,
unless they are extremely close to the line of sight; we find that
including halos fainter than $i=24$ does not improve the reconstruction. 
\Fref{fig:magcut} shows the scatter on $\kappah-\kappax$ (where $\kappah$ is
shifted so that its mean is zero, \phil{to emulate the effect of the eventual
calibration}) as a function of reconstruction magnitude limit. We do not
expect a deeper survey to decrease the size of the uncertainty in mapping
$\kappah$ onto $\kappax$. Halos at all redshifts out to the source redshift
(1.4 in this work. The redshift of the source in B1608+656) contribute to the convergence, but the largest contribution comes
from halos with $z \sim z_{\rm source}/2$.

\section{Testing the Halo Model Reconstruction on Mock Galaxy Catalogues}
\label{sec:obsMstar+z}

We now move on to consider the halo model reconstruction of line-of-sight mass
distributions given noisy astronomical observables. A typical imaging survey
can be expected to provide measurements of the positions and magnitudes of
galaxies in a field;  spectra for some of the objects may either come from a
synergistic survey, or from targeted follow-up. In this section we quantify
the uncertainties induced by inferring the halo mass and redshift from these
observables. 



As described in \Sref{sec:model:halos}, in this work we attempt to  infer halo
masses from measurements of stellar mass. We will investigate two main sources
of uncertainty: the stellar masses themselves, and placing halos at
photometric redshifts. Much work has already focused on using photometric
colours to infer stellar mass \citep[\eg][]{AugerEtal2009} and redshifts
\citep[\eg][]{BPZ}: we will estimate the likely uncertainties on stellar mass
and redshift based on this work.


\subsection{Making Mock Observational Catalogues}

\phil{While the \MS catalogues do already contain stellar masses for each
galaxy, we do not use them for two reasons. The first is that at the low mass
end, the dark matter-only \MS satellite galaxy halos are highly  stripped
relative to what we might expect in a universe containing baryons, leading to
a mismatch between the observed and simulated stellar mass--halo mass
relations.  The second is that we wanted to be able to perform the functional
test of trying to recover the convergence having assumed the correct stellar
mass--halo mass relation.} For these reasons, we assign a new true stellar
mass to each halo in the \MS catalogues, according to the empirical stellar
mass--halo mass relation of \citet{BehrooziEtal2010}. \new{When assigning stellar
masses we have treated satellites in the same way as central halos; in the real
universe both the density profile and the stellar mass--halo mass relation are
likely different for satellites and centrals, but our simple model does not include
these effects}. From these we simulate
observed stellar masses by drawing samples from  $\pr(\log{M^{*}_{\rm
obs}}|\log{M^{*}_{\rm true}})$ which we take to be a Gaussian of width
$\sigma_{M_*}$ centred on $\log(M*_{\mathrm {true}})$. Where a spectroscopic
redshift exists, stellar masses can be estimated with typical uncertainties of
0.15 dex \citep{AugerEtal2009}; however with photometric redshifts stellar
mass uncertainties are typically three times as large. We use
$\sigma_{M_*}=0.15$ dex for halos with a spectroscopic redshift and
$\sigma_{M_*}=0.45$ dex otherwise. For photometric redshift uncertainties we
draw a redshift from $\pr(z_{\rm true}|z_{\rm obs})$ which we take to be a
Gaussian of width $0.1(1+z_{\rm spec})$ centred on $z_{\rm spec}$, where
$z_{\rm spec}$ is the halo's true redshift in the \MS catalogue. \new{In our mock catalogues 
we have used the galaxy position from the \MS; this is not necessarily coincident with the
centre of the galaxy's host dark matter halo.}

\comment{The following is relevant to all sections on inferring $\kappax$
given imperfect data, so it belongs here and not in the spectroscopic redshift
first subsection.}

Given an uncertain stellar mass and redshift it is possible to infer a
halo mass using the stellar mass--halo mass relation of
\citet{BehrooziEtal2010}. This \proceedure requires an inversion of the
relation given in \citet{BehrooziEtal2010} and correctly inverting the
relation's uncertainties requires care: the \proceedure we use to do this
is given in \Aref{appendix:MSMH}. By drawing sample halo masses
and redshifts, we can infer a sample $\kappah$ using the \proceedure of
\Sref{sec:model}. Repeatedly drawing samples allows us to
estimate $\pr(\kappah|\data)$ for each reconstructed line of sight. We then
transform this into our target PDF $\pr(\kappax|\data)$ as described above.


\subsection{Reconstructing $\kappax$ given a spectroscopic redshift for every object}

\begin{figure}
\includegraphics[width=\columnwidth]{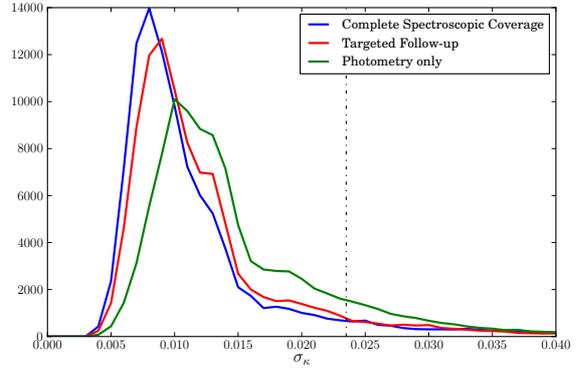}
\caption{The widths of the \infered PDFs $\pr(\kappax|\data)$ for
$10^5$ lines of sight, given different quality of data. 
Blue: spectroscopic redshift for every halo with $i<26$; 
Red: spectroscopic redshift for every halo with $i<23$, 
and every halo with $i<24$ within 1 arcminute, while all other objects just
have photometric redshifts;
Green: all the objects in the field have only photometric redshifts. The
vertical dot-dashed line marks the width of the global $\pr(\kappax)$ for all
lines of sight.}
\label{fig:reconwidths}
\end{figure}

\begin{figure}
\includegraphics[width=\columnwidth]{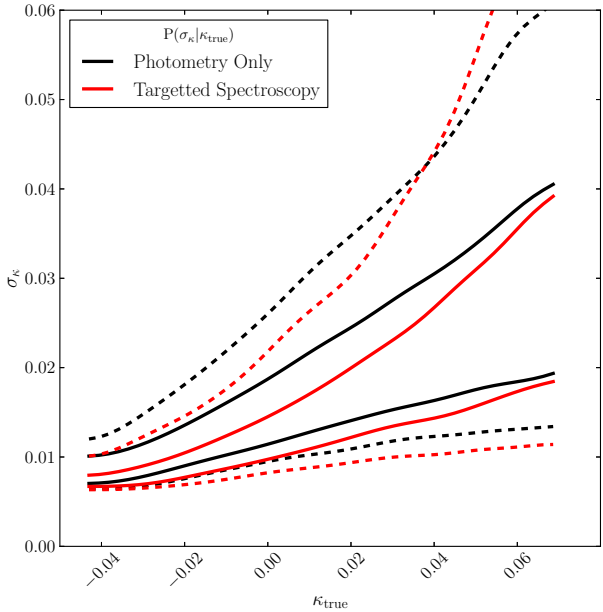}
\caption{Width of the \infered PDF $\pr(\kappax|\data)$ as a function of 
the true convergence, $\kappaxtrue$, for two different data qualities. 
Black assumes only photometric redshifts for all objects,
while red assumes a campaign of targeted spectroscopy. The region between the
solid (dashed) lines contains 68\% (95\%) of lines-of-sight.}
\label{fig:widthsvsH}
\end{figure}

The best possible reconstructions of $\kappax$ will come from having a
spectroscopic redshift for every single object in each field.
\Fref{fig:reconwidths} shows the distribution of $\kappax$ posterior 
PDF widths for various data qualities: complete spectroscopic coverage results
in the smallest widths. In terms of telescope time, such a reconstruction
would likely be prohibitively expensive, but we investigate this scenario as
an ideal case. 

\comment{PJM: The following does not belong here because \Fref{fig:widthsvsH}
does not show results for spectroscopic redshifts only!}

Reconstructing the lines of sight with perfect knowledge of the
redshift but an uncertain stellar mass, we find that the width of
$\pr(\kappax|\data)$ grows with the expectation of $\kappax$ and the ray-traced $\kappatrue$
(\Fref{fig:widthsvsH}). The low $\kappax$ lines of sight are relatively empty,
and so there are few opportunities for uncertainties in the halo masses to
\propogate into $\kappax$ uncertainties; low $\kappax$ lines of sight can be reconstructed
more precisely than high $\kappax$ lines of sight.


\subsection{Reconstructing $\kappax$ from photometry alone}

Inferring the stellar mass of a galaxy from its magnitude and colours requires
an estimate of how far away the galaxy is; without a spectroscopic redshift
the \infered stellar mass is less precise. However, obtaining photometry has
much lower observational cost; all upcoming large area photometric surveys
will reach 24th magnitude, providing sufficient data to reconstruct lines of
sight  {\it without} additional observations. In principle the photometric
redshift is correlated with the \infered stellar mass, however we do not model
this effect since the convergence from the outskirts of an individual halo is
only weakly \dependant on redshift at fixed mass due to the breadth of the
lensing kernel: the redshift uncertainty has only a small effect on the
inferred $\pr(\kappax|\data)$ in comparison to that of  the uncertain stellar
masses.  With only photometric redshifts, the uncertainty on $\kappah$ is
larger than in the spectroscopic case, and this \propogates into a broader
$\pr(\kappax|\data)$, as can be seen in \Fref{fig:reconwidths}. However, the
photometric reconstruction still typically produces a 50\% improvement
compared to the precision of the global $\pr(\kappax)$. 

\tom{With photometric data alone, $\kappah^{\rm med}$ can shift by $\sim$0.01
relative to $\kappah^{\rm med}$ given spectroscopic coverage. Since  the
$\kappax$ contribution of voids cannot change with data quality, the shift in
$\kappah^{\rm med}$ must be due to the asymmetric propagation of stellar mass
uncertainties into $\kappax$ uncertainties. As the stellar mass uncertainty
increases, $\kappah^{\rm med}$ is pushed higher.  To correctly calibrate 
$\pr(\kappax|\kappah^{\rm med},\data)$ one {\it must} include the data quality
used to generate $\kappah^{\rm med}$.}

We find that the width of $\pr(\kappax|\data)$ grows with the expectation
value of $\kappax$; this is shown in \Fref{fig:widthsvsH}. The low $\kappax$
lines of sight are relatively empty, and so there are few opportunities for
uncertainties in the halo masses to \propogate into $\kappax$ uncertainties.

%


\subsection{Reconstructing $\kappax$ with partial spectroscopic coverage}
\label{sec:obsMstar+z:targetedspec}

While a fully photometric reconstruction provides useful constraints on
$\pr(\kappax|\data)$, targeted spectroscopy can provide additional constraints
on the masses and redshifts of halos whose $\kappah$ contributions have the
largest absolute uncertainty. Obtaining spectra of bright ($i<22$) objects is
relatively fast with large telescopes: however, we find that if spectroscopic
redshifts were known for all $i<22$ galaxies in our fields the reconstruction
improves only slightly over the purely photometric reconstruction,  although
the improvement can depend strongly on the details of the particular line of
sight. Obtaining spectra for fainter objects would be correspondingly more
expensive, but if spectroscopic redshifts could be obtained for all $i<23$
galaxies (as part of a futuristic baryon acoustic oscillation survey, for
example) and all $i<24$ galaxies within 1 arcminute of the line of sight, then
we find that the $\pr(\kappax|\data)$ would be almost as precise as that from
having complete spectroscopic coverage (\Fref{fig:reconwidths}). If any object is
extremely close to the lines of sight our approximation of weakly lensing
mass-sheets will break down; also, neglecting baryonic effects will also be a poor
approximation; a spectrum will be needed
to adequately model systems with very well aligned pertrubers.
Consistent with our previous findings, the lowest $\kappax$ lines of
sight have the most constrained $\kappax$ PDFs.


\section{Systematic Errors due to Sample Selection}
\label{sec:biases}

\begin{figure*}
\includegraphics[width=\textwidth]{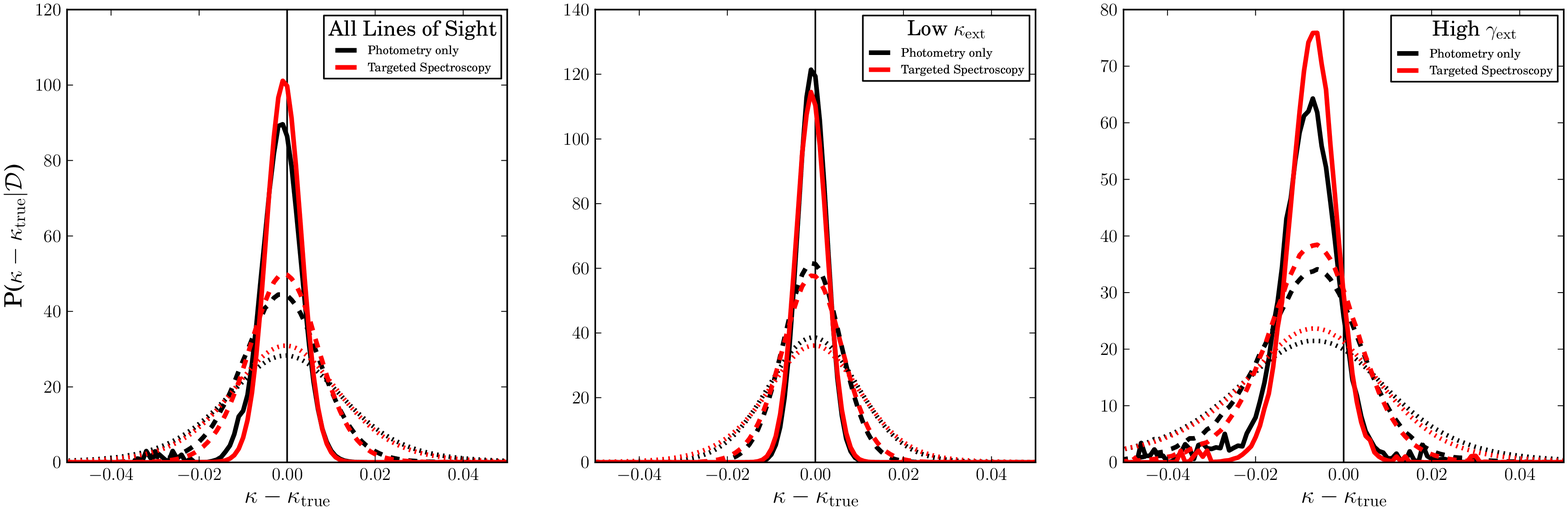}
\caption{Accuracy in $\kappax$ from combining samples of lenses.  These plots
show the expectation value of $\prod_{i=1}^N \pr_i(\kappax-\kappaxtrue|\data)$ --
deviations from zero represent biases. The solid, dashed and dotted lines
correspond to combinations of  20, 5 and 2 sightlines respectively. Black
lines show the results \infered from photometry alone, whilst red lines show
the results  from the same targeted spectroscopy campaign described in 
\Sref{sec:obsMstar+z:targetedspec}. The three panels correspond to samples of
sightlines selected in different ways. Left: randomly selected lines of
sight.  Centre: sightlines randomly selected from the 33\% of lines whose
P($\kappax$) is most tightly constrained by our model.  Right: lines of sight
with external shear of 0.05 or greater.}
\label{fig:biasplots}
\end{figure*}

While it is good to reconstruct $\kappax$ precisely, it is more important that
$\kappax$ is reconstructed accurately. The reconstruction must not be biased. 
We quantify the accuracy of our reconstructed $\pr(\kappax|\data)$ in the
following way. By shifting the \infered $\pr(\kappax|\data)$ for each line of
sight by the true (ray-traced) convergence, we obtain a  PDF that should be
centred on zero; these offset PDFs can be multiplied together for multiple
lines of sight to emulate a joint likelihood analysis, and then test for
possible bias in it:
\be
\label{eq:bias}
\mathcal{P}_N = \prod_{i=1}^N \pr_i(\kappax-\kappax^{\rm true}|\data)
\ee
In this section, we quantify the bias as the size of the deviation in the
expectation value of $\mathcal{P}_N$ from zero. If the bias of $\mathcal{P}_N$
is smaller than the half-width of $\mathcal{P}_N$, our $\kappax$ inferences
can be considered accurate. Since the width of $\mathcal{P}_N$ decreases as
more sightlines are combined, a small bias will always 
eventually be found. In the context of measuring time-delay distance at the 
one percent level, the expectation value of $|\mathcal{P}_N|$ needs to be 
significantly less than 0.01. \tom{External convergence is not the only source
of uncertainty in measuring time-delay distance. Since the statistical
distance uncertainties due to lens modelling and time delay estimation are
likely to be at the 3-4\% level \citep{SuyuEtal2010}, convolving
$\mathcal{P}_N$ with a Gaussian of width $0.04/\sqrt{N}$ is a reasonable
approximation for the likely final uncertainty on time-delay distance.}


The source of systematic error that we investigate in this paper  is that of
sample selection: we define three different example selections of lines of
sight, and compute the bias  \phil{that would result (if left unaccounted
for)} in each one in turn, as a function of data quality. The results
described below are illustrated in \Fref{fig:biasplots}.


\subsection{Selecting random lines of sight}
\label{sec:bias:random}

We first consider a purely random selection of lines of sight. With the purely
photometric reconstruction, there is no evidence of bias, even after combining
20 lines of sight. \tom{This result is a sanity check: if we had chosen lines
of sight with identical $\kappah^{\rm med}$ rather than $\kappah^{\rm
med}\pm0.003$, randomly selected lines of sight would have zero bias by
construction}. If lines of sight hosting time delay lenses are truly random, 
our method allows $\kappax$ to be reconstructed without inducing a bias (under
the basic assumptions that the universe is like the calibration simulation,
and the correct stellar-mass to halo-mass relation has been used).


\subsection{Selecting only the lines of sight with narrow $\pr(\kappax|\data)$}
\label{sec:bias:tightPDF}

Since follow-up campaigns (such as high resolution imaging, or time
variability monitoring) are expensive, it is likely that only a subset of
detected objects will be observed further. By pre-selecting objects that have 
the best-constrained convergence, the cosmological value per lens can be
increased. However, this selection \phil{(like any other)}
has the potential to induce a bias. It is
likely that only photometry will be available at the time of sub-sample
selection (although spectroscopic follow-up might be conducted at a later
date). We mimic such a selection by drawing lines of sight only from those in
the lowest third of $\sigma_{\kappa}$, given a purely photometric
reconstruction of their fields. 

We find that selecting the most tightly constrained lines of sight in this way
does not introduce a significant bias:  our reconstruction is most accurate
for the lowest $\kappax$ lines of sight. Combining 20 such systems, the bias
is at the 0.001 level.

Photometry of the field seems to be sufficient to adequately select these
empty lines of sight.  However, 
since lenses are typically massive galaxies and hence found in locally over-dense
environments, it is unlikely that  an almost empty line of sight would
actually  contain a lens; \phil{what we can say is that if the local
environment can be identified spectroscopically, and then the line
of sight mass distribution reconstructed using our method, selecting lenses
with low, well-constrained values of $\kappax$ will lead to an increase in
distance precision at no cost in bias.}


\subsection{Selecting only high shear lines of sight}
\label{sec:bias:tightPDF}

Time delay lenses are often selected for follow-up based on their images'
brightness (to make monitoring observations less expensive), their images'
separation (to decrease the covariance between the ground-based image
light-curves), or the fact that they are quadruply-imaging rather than
doubly-imaging (since this yields 3 time delays instead of 1, a higher
magnification and a more informative Einstein Ring system). All three of these
properties favour lenses with high convergence and shear along their lines of
sight. Focusing on the quad selection, we might expect many of these systems
to have significant external shear (since the cross-section for 4 image
production is so sensitive to ellipticity in the lens mass distribution (or
equivalently, external shear due to the lens' environment). 

\phil{We can test the impact of such a shear selection by selecting only lines
of sight with an external shear above some threshold.  We define a somewhat
extreme selection  of $\gammax>0.05$, and find that in this sample, $\kappax$
would be systematically underestimated at the $\sim$0.008 level (corresponding
to a 0.8\% systematic error in distance).}  In the absence of other sources of
uncertainty, this systematic error would be significant in a sample of just 5
lenses; when including other time-delay distance uncertainties a 0.8\% bias would be
significant for $\sim$20 lenses. This systematic error is mitigated if only the lines of sight with both
$\gammax>0.05$ and well constrained $\pr(\kappax|\data)$ are used, but in this
case a bias at the $\sim$0.005 level remains. 

While $\gammax>0.05$ is likely to be a much stronger selection than would
occur in reality, it is  nevertheless worth noting from this example that the
reconstruction \proceedure \phil{can be biased if lens selection functions are
extreme and unaccounted for.} Since our model does not include shear
constraints it is not surprising that a selection function based on shear can
induce a bias. A more sophisticated model that includes the shear recovered
from the lens modelling might be less susceptible to this bias.

The halo model can also be used to estimate the external shear along a line of
sight: shear is an observable that can be extracted from strong lens
modelling. However, there is a degeneracy between internal and external shear.
When the Einstein Ring imaging data are very good it is possible to
disentangle external and internal shear \citep[\eg][]{SuyuEtal2010}, but there
are still significant uncertainties. \citet{WongEtal2011} attempted to match
the shear from strong lens models with a reconstruction of the local lens
group environment, but found a tension between the strong lens model and the
reconstruction of the environment. Given the \citet{WongEtal2011} results, it
is unclear whether the external shear from lens models can be reconciled with
a line-of-sight reconstruction. Alternatively, it may be possible  to infer
external shear using weak lensing information from near the line of sight.  If
the true external shear can be measured, it provides an additional constraint
on which of the \MS lines of sight are similar to the reconstructed line of
sight. \citet{SuyuEtal2012} found that in the case of 
RXJ1131-1231, combining shear constraints
with galaxy number count over-density gave a significantly different
$\pr(\kappax|\gamma,N_{45})$ compared to the PDF from number count
over-density alone, $\pr(\kappax|N_{45})$.

Extending our model to include shear we find that given
perfect knowledge of the halo mass and redshift the ray-traced external shear
$\gamma$ and the reconstructed external shear $\gamma_{\mathrm{h}}$  are
similar, with 68 percent of lines obeying
\be
\label{eq:shearineq}
|{\pmb{\gamma}_{\mathrm{ext}}-\pmb{\gamma}_{\mathrm{h}}}| < \mathrm{0.025}
\ee 
Future work should investigate whether $\gamma_{\mathrm{h}}$ can be used
to improve the accuracy and precision of $\kappax$ estimation, given a
reconstruction of the line of sight. 


\section{Systematic Errors due to Assuming an Incorrect 
Stellar Mass--Halo Mass Relation}
\label{sec:SHAMfail}

\begin{figure*}
\includegraphics[width=\textwidth]{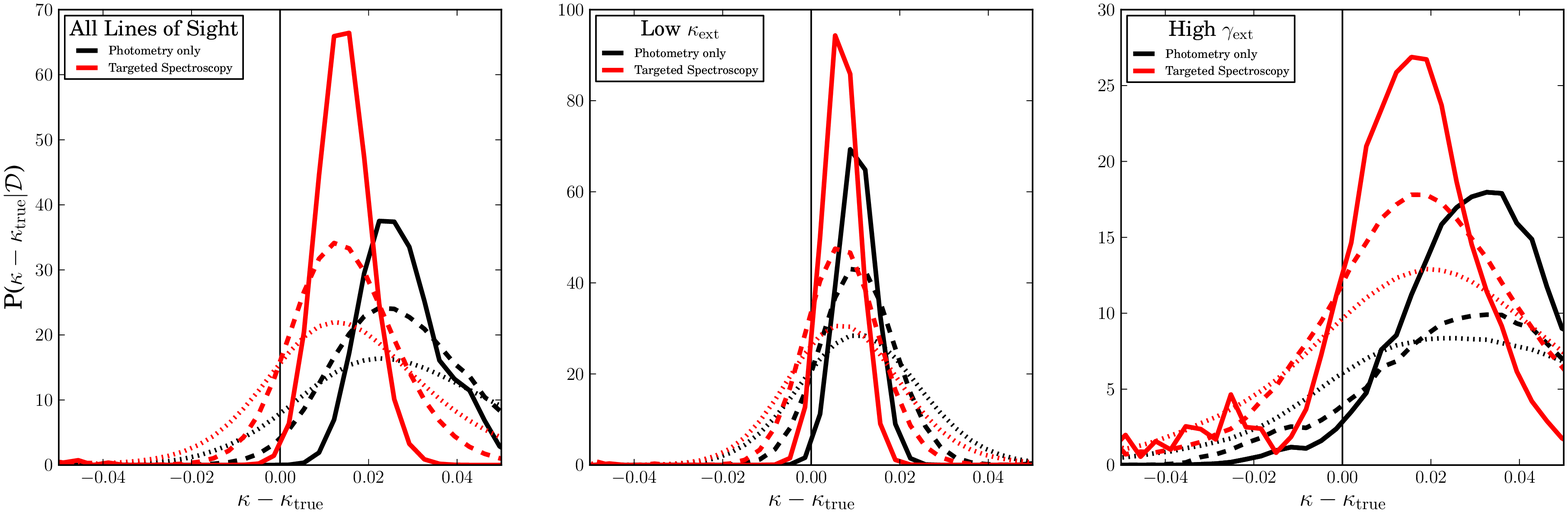}
\caption{Systematic bias in $\kappax$ due to reconstructing lines of sight with the wrong stellar mass--halo mass relation. The real stellar masses were created using the relation of \citet{MosterEtal2010}, but the reconstruction and calibration assumes the relation of \citet{BehrooziEtal2010}.  These plots
show the expectation value of $\prod_{i=1}^N \pr_i(\kappax-\kappaxtrue|\data)$ --
deviations from zero represent biases. The solid, dashed and dotted lines
correspond to combinations of  20, 5 and 2 sightlines respectively. Black
lines show the results \infered from photometry alone, whilst red lines show
the results  from the same targeted spectroscopy campaign described in 
\Sref{sec:obsMstar+z:targetedspec}. The three panels correspond to samples of
sightlines selected in different ways. Left: randomly selected lines of
sight.  Centre: sightlines randomly selected from the 33\% of lines whose
P($\kappax$) is most tightly constrained by our model.  Right: lines of sight
with external shear of 0.05 or greater.}
\label{fig:SHAMbias}
\end{figure*}

\new{

Throughout this work we have assumed that the universe's halos are populated
with galaxies whose stellar masses are determined purely by the
\citet{BehrooziEtal2010} Stellar Mass--Halo Mass relation, which we then use
in our reconstruction. In practice, the true Stellar Mass--Halo Mass relation
may well be different to the one we assume in our inference, and if it is, 
a systematic error could be incurred. It is hard to test the size of this systematic error, since we do not know how much the real universe differs from the chosen Stellar Mass--Halo Mass relation. As wider and deeper surveys are conducted more data will become available with which to construct the Stellar Mass--Halo Mass relation; this should drive the inferred Stellar Mass--Halo Mass relation closer to the truth. We are only interested in testing the effect of changing the Stellar Mass--Halo Mass relation in a way that is consistent with observational constraints; improving the observational constraints on the Stellar Mass--Halo Mass relation will decrease the the potential size of the systematic error on reconstructed $\kappax$ induced by assuming a specific Stellar Mass--Halo Mass relation.

To estimate the potential size of this systematic error, we repeat the
analysis of the previous section, still using the  the
Behroozi et al. relation to generate the stellar masses of our
calibration lines of sight and hence carry out the inference, but now using
simulated catalogue datasets that were generated from the stellar mass--halo
mass relation of Moster et al. 

This relation provides a comparable fit to the stellar mass function to the
Behroozi et al relation, so represents a plausible alternative to our assumed
form. For halo masses of $\sim 10^{12} \Msun$, both stellar mass--halo mass
relations predict a stellar mass of $\sim 10^{10.5} \Msun$, but for halos more
massive than this the stellar masses generated from the Moster relation are
systematically higher than those generated from the Behroozi relation. At halo
masses of $\sim 10^{14} \Msun$ the predicted stellar masses differ by some 
$0.25$ dex. 

Applying our Behroozi--based reconstruction to lines of sight with Moster
stellar masses results in systematic overestimates of $\kappax$. In
\Fref{fig:SHAMbias} we show this bias emerging after combining several lines
of sight. With a purely photometric reconstruction there is a typical
systematic bias of $\sim 0.025$ on the reconstructed convergence. This can be
shrunk to $\sim 0.01$ if only the low $\kappax$ lines of sight are used;
however for the high shear lines of sight the bias is $\sim 0.04$. It is not
too surprising that the low $\kappax$ lines of sight are least affected by
changes in the stellar mass--halo mass relation. Low $\kappax$ lines of sight
are relatively empty, their $\kappah$ values are small, and they do not depend
as strongly on the stellar mass--halo mass relation. In contrast, the high
shear lines of sight have significantly overestimated $\kappax$ values, since
they tend to lie close to massive halos -- the regime where the Moster
and Behroozi relations are most different. Interestingly, overestimating the
$\kappah$ values of high shear lines of sight pushes them into the region
where the $\kappah$ to $\kappax$ calibration is least certain; this 
significantly decreases the precision of the reconstructed $\kappax$. 

Despite these systematic errors, we find that the photometric reconstruction
is still sufficient to choose the best lines of sight for spectroscopic
follow-up. With targeted spectroscopy the systematic error decreases to $\sim
0.014$ for the ensemble, $\sim 0.005$ for the low kappa sample and $\sim
0.018$ for the high shear sample. These results provides further motivation for
prioritizing the lenses that reside on the under-dense lines of sight.

The systematic errors reported here are perhaps overly pessimistic, because
the bias is due to differences in the stellar masses predicted for high mass
halos. If new observations can discriminate between the high mass end of the
Behroozi et al. and Moster et al. relations, then the
systematic error on the reconstruction will decrease. Similarly, incorporating
additional information about the high mass group and cluster-scale halo
systems, such as their richness or other occupation statistics, would help
mitigate the stellar to halo mass conversion error. }


\section{Discussion}
\label{sec:discuss}


The total convergence along a line of sight is strongly correlated with the
reconstructed $\kappah$. However, since our model ignores voids and assumes
all halos follow a spherical truncated-NFW profile our halo model does not
include all of the relevant physics; hence, the width of our resulting
$\pr(\kappax|\data)$ is still typically $\sim$0.01 for any given lightcone,
even assuming perfect knowledge of every halo's virial mass and redshift. To
make further progress a more advanced treatment of both voids and halos will
be needed. \citet{KarpenkaEtal2012} find that their supernova data prefer, in
the context of their simple halo model, a truncated isothermal profile for the
galaxy mass distributions. Galaxy-galaxy and cluster-galaxy  lensing studies
\citep[\eg][]{GavazziEtal2007,JohnstonEtal2007, LagattutaEtal2010} also suggest an approximately
isothermal profile for the mass-shear correlation function; however, at large
radii these profiles are dominated by the effects of neighbouring halos, which
are already taken into account by our model.  Further iterations on the
analysis presented here should include the stellar mass distributions that
cause the isothermal density profiles in the centres of galaxies, although we
do not expect this to have a big impact on the reconstructions. Likewise, 
halo ellipticity
may have some small effect on the predicted convergence. It
is possible that baryonic physics may alter the radial profile of dark
matter halos, even on large scales. \citet{Semboloni+2012} found that baryonic
physics had a 30-40\% impact on three-point shear statistics, it is unclear
how baryons affect one-point convergence statistics compared to the pure dark
matter simulations used in this work.

Clusters, filaments and voids are more difficult to model, but simple forms
could in principle be derived from the structures found in the simulations,
and included in the halo model in some way.  Most importantly for the highest
$\kappax$ lines of sight will be a better group and cluster model:
\citet{MomchevaEtal2006} and  \citet{WongEtal2011} recommend using models
where dark mass is assigned to both galaxy-scale halos and halos for the
groups and clusters they occupy. To make this work, one could incorporate a
group and cluster finder, and then infer halo mass from optical richness
\citep[\eg][]{MaxBCG} in tandem with BCG stellar mass. \new{A group finder would also
allow us to identify central and satellites galaxies; in this work we have neglected
the differences between central halos and satellites. However, given perfect knowledge of halo masses
we found that only in 5 percent of sightlines do satellites
contribute over 20 percent of the sightline's total $\kappax$. Introducing a group finder
is unlikely to provide major improvement to the reconstruction of satellites.}
\tom{Meanwhile,
improving the modelling of voids could be done using suggestions by
\citet{voids2} or \citet{voids1} combined with a void finder such as in
\citet{voids3}. Combining an advanced halo model with a void model will allow
a direct test of the halo approximation: using real photometric data one could
reconstruct lines of sight and compare against measured weak lensing shears.
This would ensure that the reconstruction model reproduces the contents of the
real universe, rather than a particular simulation.}

The reduction of $\pr(\kappah|\data)$ to just its median value, before using
it to look up the appropriate $\pr(\kappax)$ from the simulated sightline
ensemble, is reminiscent of the procedure followed by \citet{SuyuEtal2010},
and \citet{Greene}. The reconstruction method presented here can be
seen as the limiting case of the weightings explored by Greene et al: rather
than using weighted number counts in an aperture, we treat each galaxy
individually, and predict $\kappa_{\rm h,obs}^{\rm med}$ for use in their
place. Given photometry alone, our reconstruction represents a $\sim$50\% improvement of
the precision compared the number counts used in \citet{SuyuEtal2010}. Greene et al.
made important progress for the most over-dense lines of sight; our
reconstructions are $\sim$30\% more precise than those of Greene et al.
The use of relative over-densities by Suyu et al. and Greene et al. may decrease 
their sensitivity to the choice of calibration simulation; this work's use of
absolute convergence may be less robust to changes in the calibration simulation. 


In this work, we have used  the \MS as a calibration tool. If the real
Universe is not like the \MS, then this calibration will be incorrect.
Repeating this study using a different simulation to make the test catalogues
would allow us to quantify the sensitivity of our results to the simulation
used. Similarly the use of a different semi-analytic model to paint galaxies
onto halos would enable similar exploration of these systematic errors. 
\citet{SuyuEtal2010}
used the relative over-density to mitigate the effect of simulation and
simulation, but it does not completely remove the dependence on simulation.
\new{We have tested the effect of using a different stellar mass--halo mass relation
and found that this can induce significant systematic biases if the true relation is unknown.} Future work should include
investigating the robustness of the reconstruction to these systematic errors \new{and develop prescriptions that mitigate against them}


\section{Conclusions}
\label{sec:conclude}

In this work we have investigated a simple halo model prescription for
reconstructing all the mass along a line of sight to an intermediate redshift
source. We have used the ray-traced lensing convergence along lines of sight
through the Millennium Simulation to calibrate estimates of the total
convergence made by summing the convergences due to each object in a
photometric catalogue. Having found that the reconstruction process is
accurate given this calibration and perfect knowledge of halo mass and
redshift, we investigated the effects of reasonable uncertainties in the
stellar mass and redshift of each halo, and propagated these uncertainties
into a $\pr(\kappax|\data)$ for each line of sight. We also defined three
different possible line of sight selections, and investigated the possible 
bias induced by these selections. We draw the following conclusions:

\begin{itemize} 

\item Despite our model's simplicity, its reconstructed $\kappah$ values are
good tracers of $\kappax$. $\kappah$ is biased due to our ignorance of voids,
groups and clusters, and the unseen mass in small halos and filaments, but we
find that this can be calibrated out using the Millennium Simulation
catalogues. We found that with perfect knowledge of every halo's mass and
redshift,  calibrated reconstruction of a typical line of sight  gives an
unbiased estimate of $\kappax$ that is not perfect, but is $1.8$ 
times more precise than the global
$\pr(\kappax)$. This factor quantifies the limit to which the halo model can
represent the external convergence due to line of sight mass structure.

\item With uncertain halo masses and redshifts, we find that
$\pr(\kappah|\data)$ can still be calibrated in order to infer
$\pr(\kappax|\data)$ from the ensemble of simulated lines of sight, but the
resulting PDFs tend to be much broader than for perfect halo mass and redshift
reconstructions.

\item It is very rare for halos further than 2 arcminutes from the line of
sight to make a significant contribution to $\pr(\kappax|\data)$; likewise, 
including
halos whose host galaxy is less luminous than $i=24$ does not significantly
improve our reconstructions.  A photometric survey to this depth of a
4$\times$4 arcminute patch around the target would approach the limiting
uncertainties of our simple reconstruction recipe, and yield a 
$\pr(\kappax|\data)$ that has, on average, a width of 0.016 (inducing a $\sim$1.6\%
uncertainty on time-delay distances). This is 1.5 times
less broad than the global $\pr(\kappax)$. 

\item  For the most over-dense lines of sight, the reconstruction produces a
broad PDF; conversely, we find that the lines of sight with the sharpest
inferred $\pr(\kappax|\data)$ are typically under-dense. Since the
reconstructions described here can be performed before follow-up time is
invested, it will be possible to select targets by the precision of their
$\kappax$ estimates. This will be useful in an era when there are more known
lenses than can be followed-up. Photometric data are sufficient to select the best lines of sight for
follow-up. Spectroscopic coverage helps to tighten  $\pr(\kappax|\data)$ for
most lines of sight.

\item Selecting from the highest precision 33\% of the reconstructed lines of
sight results in a sample with low bias, $\sim$0.1\% in terms of the
time-delay distance. \tom{For these lines of sight the reconstructed
$\pr(\kappax|\data)$ induces a statistical uncertainty of $\sim$1\% on
individual time-delay distances.}

\item Conversely, selecting lines of sight with high shear ($|\gamma| > 0.05$; 
a somewhat extreme selection, but one that  could potentially arise from 
focusing on four-image or wide separation lenses) results in a sample with
bias of 0.8\% in time delay distance (and hence $H_0$). Including realistic estimates of other
sources of uncertainty a systematic error of 0.8\% would become significant in
a sample of $\sim$20 lenses. The addition of shear constraints to our model might
alleviate this potential systematic.

\item \new{Reconstructing a line of sight with an incorrect stellar mass--halo mass relation can introduce systematic errors on the inferred time-delay distance at the percent level. This error can be reduced by tightening the observational limits on the stellar mass--halo mass relation at the high mass end.}


\end{itemize}




While we have tested our reconstruction method against realistic simulations
of line of sight mass distributions in the Universe, our method is also 
calibrated to these simulations. Testing this assumption in the short term,
and replacing it with an empirical calibration (or hierarchical inference) in
the longer term, are worthy goals for future work. The halo model framework is
flexible enough to enable many such improvements, including, quite naturally,
the incorporation of more information from observations.


\section*{Acknowledgements}
TEC thanks Vasily Belokurov for supervision, guidance and suggestions.
We thank Risa Wechsler and Peter Behroozi 
for useful discussions and suggestions. We are grateful to the referee
for suggesting improvements to the original manuscript.
TEC acknowledges support from STFC in the form of a research studentship.
PJM was given support by the Kavli Foundation and the Royal 
Society, in the form of research fellowships.
%
%
SH and RDB, and CDF, acknowledge support by the National Science Foundation
(NSF), grant numbers AST-0807458 and  AST-0909119, respectively.
SHS, ZG  and TT gratefully acknowledge support from the Packard Foundation in
the form of a Packard Research Fellowship to TT and from the 
NSF grant AST-0642621.
LVEK acknowledges the support by an NWO-VIDI programme subsidy
(programme number 639.042.505).
%
%
The Millennium Simulation databases used in this paper and the web application
providing online access to them were constructed as part of the activities of
the German Astrophysical Virtual Observatory.

Our code for reconstructing lines of sight is publicly available at {https://github.com/drphilmarshall/Pangloss.}


\appendix


\section{Inferring $\Mhalo$ given a noisy measurement of $\Mstarobs$}
\label{appendix:MSMH}

To estimate the convergence caused by a halo we need to know its mass; how can 
halo mass be \infered from a noisy estimate of the stellar mass $\Mstarobs$
of a galaxy at redshift $z$. We seek the posterior
PDF $\Pr(\Mhalo|\Mstarobs,z)$, which can be expanded as follows:

\begin{eqnarray}
&& \Pr(\Mhalo|\Mstarobs,z) = \notag\\
&& \int d\Mstar \Pr(\Mhalo|\Mstar,z) \Pr(\Mstar|\Mstarobs,z), \notag\\
&\propto& \int d\Mstar \Pr(\Mstarobs|\Mstar) \Pr(\Mstar|\Mhalo,z) \Pr(\Mhalo|z),
\label{eq:mhalo-mstar}
\end{eqnarray}
where we have used Bayes' Theorem twice to replace
$\Pr(\Mhalo|\Mstar,z) \Pr(\Mstar|z)$ with 
$\Pr(\Mstar|\Mhalo,z) \Pr(\Mhalo|z)$, and 
to invert $\Pr(\Mstar|\Mstarobs)$ into the sampling
distribution $\Pr(\Mstar|\Mstarobs)$, which we recognise as the likelihood
function for the observed stellar mass. Note that the ``true'' $\Mstar$ of the
galaxy is marginalised out: we are only interested in inferring the halo
mass. The last two terms in
\eqref{eq:mhalo-mstar} are the $\Mstar-\Mhalo$ relation from
\citet{BehrooziEtal2010}, and the halo mass function $\Pr(\Mhalo|z)$, at the
given redshift. We can
tabulate the product of these two from our Millennium Simulation catalogue,
constructing a two-dimensional histogram of halo masses and their associated
true stellar masses (drawn from the Behroozi relation). 

For each galaxy, we compute the likelihood function for its $\Mstarobs$ as a
function of the unknown $\Mstar$, and multiply it by our tabulated joint PDF.
This heavily downweights halos with $\Mstar$ values outside the observed
range. We then do the marginalisation integral by Monte Carlo, drawing
(two-dimensional) sample parameter vectors
from the down weighted histogram, discarding the $\Mstar$ values, and
constructing a one-dimensional histogram that is an estimate of
$\Pr(\Mhalo|\Mstarobs)$.

If the redshift of the galaxy is uncertain, we need to take this
uncertainty into account; for example, for each sample drawn from the
photometric redshift posterior PDF $\Pr(z|{\rm colors})$, we can draw a
sample $\Mhalo$ using the above procedure.





\begin{thebibliography}{99}


\bibitem[{{Auger} {et~al.}(2007){Auger}, {Fassnacht}, {Abrahamse}, {Lubin}, \&
  {Squires}}]{AugerEtal2007}
{Auger}, M.~W., {Fassnacht}, C.~D., {Abrahamse}, A.~L., {Lubin}, L.~M., \&
  {Squires}, G.~K. 2007, \aj, 134, 668

\bibitem[Auger(2008)]{Auger2008} Auger, M.~W.\ 2008, \mnras, 383, 
L40 

\bibitem[\protect\citeauthoryear{Auger et al.}{2009}]{AugerEtal2009} 
Auger M.~W., Treu T., Bolton A.~S., Gavazzi R., Koopmans L.~V.~E., Marshall 
P.~J., Bundy K., Moustakas L.~A., 2009, ApJ, 705, 1099 


\bibitem[\protect\citeauthoryear{Auger et al.}{2010}]{AugerEtal2010} 
Auger M.~W., Treu T., Bolton A.~S., Gavazzi R., Koopmans L.~V.~E., Marshall 
P.~J., Moustakas L.~A., Burles S., 2010, ApJ, 724, 511 


\bibitem[\protect\citeauthoryear{Baltz, Marshall, 
\& Oguri}{2009}]{BMO} Baltz E.~A., Marshall P., Oguri M., 2009, JCAP, 1, 15 


\bibitem[\protect\citeauthoryear{Behroozi, Conroy, 
\& Wechsler}{2010}]{BehrooziEtal2010} Behroozi P.~S., Conroy C., Wechsler R.~H., 2010, ApJ, 717, 379 


\bibitem[\protect\citeauthoryear{Ben{\'{\i}}tez}{2000}]{BPZ} Ben{\'{\i}}tez N., 2000, ApJ, 536, 571 


\bibitem[{{Brada{\v c}} {et~al.}(2009){Brada{\v c}}, {Treu}, {Applegate},
  {Gonzalez}, {Clowe}, {Forman}, {Jones}, {Marshall}, {Schneider}, \&
  {Zaritsky}}]{BradacEtal2009}
{Brada{\v c}}, M., {et~al.} 2009, \apj, 706, 1201

\bibitem[Cropper et al.(2012)]{Euclid} Cropper, M., Cole, R., 
James, A., et al.\ 2012, arXiv:1208.3369 


\bibitem[\protect\citeauthoryear{Collett et 
al.}{2012}]{CollettEtal2012a} Collett T.~E., Auger M.~W., Belokurov V., 
Marshall P.~J., Hall A.~C., 2012, MNRAS, 424, 2864 




\bibitem[{{Dalal} {et~al.}(2005){Dalal}, {Hennawi}, \& {Bode}}]{DalalEtal2005}
{Dalal}, N., {Hennawi}, J.~F., \& {Bode}, P. 2005, \apj, 622, 99


\bibitem[\protect\citeauthoryear{De Lucia 
\& Blaizot}{2007}]{DeLucia+Blaizot2007} De Lucia G., Blaizot J., 2007, MNRAS, 375, 2 


\bibitem[Fadely et al.(2010)]{FadelyEtal2009} Fadely, R., Keeton, 
C.~R., Nakajima, R., \& Bernstein, G.~M.\ 2010, \apj, 711, 246 


\bibitem[\protect\citeauthoryear{Falco, Gorenstein, 
\& Shapiro}{1985}]{FalcoEtal1985} Falco E.~E., Gorenstein M.~V., Shapiro I.~I., 1985, ApJ, 289, L1 


\bibitem[{{Fassnacht} \& {Lubin}(2002)}]{Fassnacht+Lubin2002}
{Fassnacht}, C.~D., \& {Lubin}, L.~M. 2002, \aj, 123, 627

\bibitem[Fassnacht et al.(2006)]{FassnachtEtal2006} Fassnacht, C.~D., 
Gal, R.~R., Lubin, L.~M., et al.\ 2006, \apj, 642, 30 

\bibitem[\protect\citeauthoryear{Fassnacht, Koopmans, 
\& Wong}{2011}]{FassnachtEtal2011} Fassnacht C.~D., Koopmans L.~V.~E., Wong K.~C., 2011, MNRAS, 410, 2167 

\bibitem[Foster 
\& Nelson(2009)]{voids3} Foster, C., \& Nelson, L.~A.\ 2009, \apj, 699, 1252 

\bibitem[Gavazzi et al.(2007)]{GavazziEtal2007} Gavazzi, R., Treu, T., 
Rhodes, J.~D., et al.\ 2007, \apj, 667, 176 

\bibitem[\protect\citeauthoryear{Gavazzi et 
al.}{2008}]{GavazziEtal2008} Gavazzi R., Treu T., Koopmans L.~V.~E., 
Bolton A.~S., Moustakas L.~A., Burles S., Marshall P.~J., 2008, ApJ, 677, 
1046 

\bibitem[Greene et al.(2013)]{Greene} Greene, Z.~S., Suyu, S.~H., Treu, T., et al.\ 2013, arXiv:1303.3588 

\bibitem[Gunnarsson et al.(2006)]{GunnarssonEtal2006} Gunnarsson, C., 
Dahl{\'e}n, T., Goobar, A., J{\"o}nsson, J., M{\"o}rtsell, E.\ 2006, \apj, 640, 417 

\bibitem[Higuchi et al.(2012)]{voids1} Higuchi, Y., Oguri, M., 
\& Hamana, T.\ 2012, arXiv:1211.5966 

\bibitem[{{Hilbert} {et~al.}(2007){Hilbert}, {White}, {Hartlap}, \&
  {Schneider}}]{HilbertEtal2007}
{Hilbert}, S., {White}, S.~D.~M., {Hartlap}, J., \& {Schneider}, P. 2007,
  \mnras, 382, 121

\bibitem[\protect\citeauthoryear{Hilbert et 
al.}{2009}]{HilbertEtal2009} Hilbert S., Hartlap J., White S.~D.~M., Schneider P., 2009, A\&A, 499, 31 

\bibitem[Hilbert et al.(2011)]{HilbertEtal2011} Hilbert, S., Gair, 
J.~R., \& King, L.~J.\ 2011, \mnras, 412, 1023 

\bibitem[Holder 
\& Schechter(2003)]{Holder+Schechter2003} Holder, G.~P., \& Schechter, P.~L.\ 2003, \apj, 589, 688 


\bibitem[\protect\citeauthoryear{Holz 
\& Wald}{1998}]{Holz+Wald1998} Holz D.~E., Wald R.~M., 1998, PhRvD, 58, 063501 

\bibitem[Holz 
\& Linder(2005)]{Holz+Linder2005} Holz, D.~E., \& Linder, E.~V.\ 2005, \apj, 631, 678 

\bibitem[J{\"o}nsson et al.(2010)]{JonssonEtal2010} J{\"o}nsson, J., 
Dahl{\'e}n, T., Hook, I., Goobar, A., M{\"o}rtsell, E.\ 2010, \mnras, 402, 526 

\bibitem[Johnston et al.(2007)]{JohnstonEtal2007} Johnston, D.~E., 
Sheldon, E.~S., Wechsler, R.~H., et al.\ 2007, arXiv:0709.1159 


\bibitem[Karpenka et al.(2012)]{KarpenkaEtal2012} Karpenka, N.~V., 
March, M.~C., Feroz, F., \& Hobson, M.~P.\ 2012, arXiv:1207.3708 

\bibitem[Kauffmann et al.(1999)]{KauffmannEtal1999} Kauffmann, G., 
Colberg, J.~M., Diaferio, A., \& White, S.~D.~M.\ 1999, \mnras, 303, 188
\bibitem[Krause et al.(2012)]{voids2} Krause, E., Chang, 
T.-C., Dor{\'e}, O., \& Umetsu, K.\ 2012, arXiv:1210.2446 

\bibitem[{{Keeton} \& {Zabludoff}(2004)}]{Keeton+Zabludoff2004}
{Keeton}, C.~R., \& {Zabludoff}, A.~I. 2004, \apj, 612, 660

\bibitem[\protect\citeauthoryear{Keeton 
\& Moustakas}{2009}]{Keeton+Moustakas2009} Keeton C.~R., Moustakas L.~A., 2009, ApJ, 699, 1720 


\bibitem[Koester et al.(2007)]{MaxBCG} Koester, B.~P., McKay, 
T.~A., Annis, J., et al.\ 2007, \apj, 660, 221 

\bibitem[Koopmans et al.(2003)]{KoopmansEtal2003} Koopmans, L.~V.~E., 
Treu, T., Fassnacht, C.~D., Blandford, R.~D., 
\& Surpi, G.\ 2003, \apj, 599, 70 


\bibitem[Koopmans(2004)]{Koopmans2004} Koopmans, L.~V.~E.\ 2004, 
arXiv:astro-ph/0412596 

\bibitem[Lagattuta et al.(2010)]{LagattutaEtal2010} Lagattuta, D.~J., 
Fassnacht, C.~D., Auger, M.~W., et al.\ 2010, \apj, 716, 1579 

\bibitem[LSST Science Collaboration (2009)]{LSST} LSST 
Science Collaboration: Abell, P.~A., Allison, J., et al.\ 2009, 
arXiv:0912.0201 

\bibitem[Macci{\`o} et al.(2008)]{MaccioEtal2008} Macci{\`o}, A.~V., 
Dutton, A.~A., \& van den Bosch, F.~C.\ 2008, \mnras, 391, 1940 

\bibitem[Mandelbaum et al.(2009)]{Mandelbaum} Mandelbaum, R., van 
de Ven, G., \& Keeton, C.~R.\ 2009, \mnras, 398, 635 



\bibitem[{{Momcheva} {et~al.}(2006){Momcheva}, {Williams}, {Keeton}, \&
  {Zabludoff}}]{MomchevaEtal2006}
{Momcheva}, I., {Williams}, K., {Keeton}, C., \& {Zabludoff}, A. 2006, \apj,
  641, 169


\bibitem[Moster et al.(2010)]{MosterEtal2010} Moster, B.~P., 
Somerville, R.~S., Maulbetsch, C., et al.\ 2010, \apj, 710, 903 


\bibitem[Nakajima et al.(2009)]{NakajimaEtal2009} Nakajima, R., 
Bernstein, G.~M., Fadely, R., Keeton, C.~R., 
\& Schrabback, T.\ 2009, \apj, 697, 1793 

\bibitem[\protect\citeauthoryear{Navarro, Frenk, 
\& White}{1997}]{NFW} Navarro J.~F., Frenk C.~S., White S.~D.~M., 1997, ApJ, 490, 493 


\bibitem[\protect\citeauthoryear{Neto et al.}{2007}]{Neto2007} 
Neto A.~F., et al., 2007, MNRAS, 381, 1450 


\bibitem[{{Oguri} \& {Takahashi}(2006)}]{Oguri+Takahashi2006}
{Oguri}, M., \& {Takahashi}, K. 2006, \prd, 73, 123002

\bibitem[\protect\citeauthoryear{Oguri 
\& Marshall}{2010}]{Oguri+Marshall2010} Oguri M., Marshall P.~J., 2010, MNRAS, 405, 2579 


\bibitem[\protect\citeauthoryear{Oke}{1974}]{Oke1974} Oke 
J.~B., 1974, ApJS, 27, 21 

\bibitem[Schneider(2006)]{Schneider2006} Schneider, P.\ 2006, 
in Saas-Fee Advanced Course 33: Gravitational Lensing: Strong, Weak and Micro, 
269 

\bibitem[Semboloni et al.(2012)]{Semboloni+2012} Semboloni, E., 
Hoekstra, H., \& Schaye, J.\ 2012, arXiv:1210.7303 


\bibitem[\protect\citeauthoryear{Springel et 
al.}{2005}]{SpringelEtal2005} Springel V., et al., 2005, Natur, 435, 629 


\bibitem[\protect\citeauthoryear{Suyu}{2012}]{Suyu2012a} Suyu, S.~H.\ 2012, \mnras, 426, 868 

\bibitem[\protect\citeauthoryear{Suyu et al.}{2012}]{SuyuEtal2012} 
Suyu S.~H., et al., 2012, arXiv, arXiv:1208.6010 


\bibitem[\protect\citeauthoryear{Suyu et al.}{2010}]{SuyuEtal2010} 
Suyu S.~H., Marshall P.~J., Auger M.~W., Hilbert S., Blandford R.~D., 
Koopmans L.~V.~E., Fassnacht C.~D., Treu T., 2010, ApJ, 711, 201 

\bibitem[Takahashi et al.(2011)]{TakahashiEtal2011} Takahashi, R., Oguri, 
M., Sato, M., \& Hamana, T.\ 2011, \apj, 742, 15 


\bibitem[Treu et al.(2009)]{TreuEtal2009} Treu, T., Gavazzi, R., 
Gorecki, A., et al.\ 2009, \apj, 690, 670 

\bibitem[\protect\citeauthoryear{Vale 
\& White}{2003}]{Vale+White2003} Vale C., White M., 2003, ApJ, 592, 699 

\bibitem[Wambsganss et al.(2004)]{WambsganssEtal2004} Wambsganss, J., 
Bode, P., \& Ostriker, J.~P.\ 2004, \apjl, 606, L93 

\bibitem[{{Wang} \& {Dai}(2011)}]{Wang+Dai2011}
{Wang}, F.~Y., \& {Dai}, Z.~G. 2011, \aap, 536, A96


\bibitem[{{Williams} {et~al.}(2006){Williams}, {Momcheva}, {Keeton},
  {Zabludoff}, \& {Leh{\'a}r}}]{WilliamsEtal2006}
{Williams}, K.~A., {Momcheva}, I., {Keeton}, C.~R., {Zabludoff}, A.~I., \&
  {Leh{\'a}r}, J. 2006, \apj, 646, 85


\bibitem[\protect\citeauthoryear{Wong et al.}{2011}]{WongEtal2011} 
Wong K.~C., Keeton C.~R., Williams K.~A., Momcheva I.~G., Zabludoff A.~I., 
2011, ApJ, 726, 84 


\bibitem[{{Wong} {et~al.}(2012){Wong}, {Ammons}, {Keeton}, \&
  {Zabludoff}}]{WongEtal2012}
{Wong}, K.~C., {Ammons}, S.~M., {Keeton}, C.~R., \& {Zabludoff}, A.~I. 2012,
  \apj, 752, 104

\bibitem[\protect\citeauthoryear{Wyithe et al.}{2011}]{WyitheEtal2011} Wyithe, J.~S.~B., Yan, 
H., Windhorst, R.~A., \& Mao, S.\ 2011, \nat, 469, 181 



\end{thebibliography}


\label{lastpage}
\bsp

\end{document}